\documentclass[12pt,preprint]{aastex}

\shorttitle{BCG Globular Cluster Systems}
\shortauthors{Harris et al.}

\begin{document}

\title{Globular Cluster Systems in
Brightest Cluster Galaxies:  Bimodal Metallicity Distributions and 
the Nature of the High-Luminosity Clusters}

\author{William E. Harris}
\affil{Department of Physics \& Astronomy, McMaster University,
  Hamilton ON L8S 4M1}
\email{harris@physics.mcmaster.ca}

\author{Bradley C. Whitmore}
\affil{Space Telescope Science Institute, 3700 San Martin Drive,
    Baltimore MD 21218}
\email{whitmore@stsci.edu}

\author{Diane Karakla}
\affil{Space Telescope Science Institute, 3700 San Martin Drive,
    Baltimore MD 21218}
\email{dkarakla@stsci.edu}

\author{Waldemar Oko\'n}
\affil{Department of Physics \& Astronomy, McMaster University,
  Hamilton ON L8S 4M1}
\email{okon@physics.mcmaster.ca}

\author{William A. Baum}
\affil{Astronomy Department, University of Washington, Seattle WA 98195}
\email{baum@astro.washington.edu}

\author{David A. Hanes}
\affil{Department of Physics, Queen's University,
    Kingston ON K7L 3N6}
\email{hanes@astro.queensu.ca}

\author{J. J. Kavelaars}
\affil{Herzberg Institute of Astrophysics, National Research Council
    of Canada, 5071 W.Saanich Road, Victoria BC V9E 2E7}
\email{jj.kavelaars@nrc-cnrc.gc.ca}

\clearpage

\begin{abstract}
We present new $(B,I)$ photometry for the globular cluster systems
in eight Brightest Cluster Galaxies (BCGs), obtained with the
ACS/WFC camera on the Hubble Space Telescope.  In the very rich
cluster systems that reside within these giant galaxies, we find
that all have strongly bimodal color distributions 
that are clearly resolved by the metallicity-sensitive
$(B-I)$ index.  Furthermore, the mean colors and internal color range of
the  blue subpopulation are remarkably similar from one galaxy
to the next, to well within the $\pm 0.02 - 0.03$-mag
uncertainties in the foreground
reddenings and photometric zeropoints.  By contrast, the mean color
and internal color range for the red subpopulation differ 
from one galaxy to the next by twice as much as the blue population.
All the BCGs show population gradients, with much higher relative numbers
of red clusters within 5 kpc of their centers, consistent with their
having formed at later times than the blue, metal-poor population.
A striking new feature of the color distributions
emerging from our data is that for the brightest clusters
($M_I < -10.5$) {\sl the color distribution becomes broad and less
obviously bimodal}.
This effect was first noticed by Ostrov et al. (1998) and
Dirsch et al. (2003) for the Fornax giant NGC 1399; our data suggest that
it may be a characteristic of many BCGs and perhaps other large galaxies.
Our data indicate that the blue (metal-poor) clusters brighter
than $M_I \simeq -10$ become progressively redder with increasing
luminosity, following a mass/metallicity scaling relation $Z \sim M^{0.55}$.  
A basically similar relation has been found for M87 by Strader et al. (2005).
We argue that these GCS characteristics 
are consistent with a hierarchical-merging galaxy formation picture
in which the metal-poor clusters formed in protogalactic clouds or dense 
starburst complexes
with gas masses in the range $10^7 - 10^{10} M_{\odot}$, but where the
more massive clusters on average formed in bigger clouds with deeper
potential wells where more pre-enrichment could occur.
\end{abstract}


\section{Introduction}

Half a century ago, \citet{baum55} first explored photometrically
the remarkably large population of globular clusters in the
giant elliptical galaxy M87.
Since then, it has become increasingly clear that giant E galaxies
exhibit by far the greatest variety of globular cluster system (GCS)
properties, particularly
in their metallicity distribution and specific frequency 
\citep[e.g.][]{h01}.
These systems thus have the potential to exert a wide range of constraints
on galaxy formation models.  During the past several years, 
multicolor photometric studies of GCSs, especially 
with HST \citep[for example,][]{whit95,neilsen99,geb99,grillmair99,kun99,kun01,larsen01} 
and with wide-field ground-based
cameras \citep[][among others]{gei96,hhg04,dirsch03,
dirsch05,rhode04}, have generated an enormous increase in the
observational data for this subject.  

Much of this recent work has concentrated
on the metallicity distribution function (MDF) of the globular clusters.
Because the broadband colors of old star clusters are, 
fortunately, sensitive to their mean heavy-element abundance,
useful MDFs can be constructed efficiently with large
statistical samples of clusters gained from deep imaging studies.
These investigations have made it 
increasingly evident that the classic, old globular clusters can 
be divided into two predominant sub-populations in most large
galaxies, differing by an order of magnitude (1 dex) in their
heavy-element enrichment.  This ``bimodality paradigm'' was 
clearly suggested a decade ago \citep{zepf93}
and has steadily been reinforced as the quality and size of the
databases have increased.

The largest GCS populations of all are found in the Brightest Cluster
Galaxies (BCGs) at the centers of rich clusters,
many of which are high-specific-frequency systems \citep{h01,hpm95,blake99}.
Because BCGs are rare, GCS photometry of high enough
quality to investigate the MDF in detail
has been produced for only the few nearest
ones, such as M87 \citep{whit95,kun99}, NGC 1399 \citep{dirsch03},
and NGC 5128 \citep{hhg04}.  But many other BCGs lie within reach of
the HST cameras, and they constitute a rich and relatively
untapped resource for GCS studies.  

A major problem in the previous photometry with HST -- driven
in turn by the low blue sensitivity of the WFPC2 camera -- is that
much of it used the $(V-I)$ color index, which is relatively
insensitive to cluster metallicity.  Since the metal-poor clusters
have mean colors $(V-I)_0 \simeq 0.9$ and the metal-rich
ones have $(V-I)_0 \simeq 1.1$, typical random measurement uncertainties
of $\pm 0.1$ mag may easily obscure the differences between the two
modes, making them difficult to extract from the observed
color histograms.  It is even more difficult to accurately extract
the intrinsic {\sl metallicity dispersions} within the blue and red modes
when these dispersions are comparable with, or less than, the photometric measurement
scatter.  In addition, the relatively
small WFPC2 field size did not allow large cluster samples to be
measured for most nearby galaxies, thus hampering the statistical 
quality with which the MDF could be defined \citep[see, e.g. the many examples
in][]{geb99,kun01,larsen01}.  Many of the ground-based studies
\citep[e.g.][]{hghh92,zepf95,gei96,dirsch03,hhg04} have circumvented
these problems by using wide-field mosaic cameras and by employing much
more sensitive color indices, particular the Washington $(C-T_1)$ index.

The Advanced Camera for Surveys (ACS) on HST provides a gain
in technical capability of nearly an order of magnitude over
WFPC2 for programs of this type:  it has a higher
overall sensitivity, its Wide Field Channel mode
has a field of view three times larger in area, and it has an 
increased blue sensitivity which allows use of color indices such as $(B-I)$
or $(g'-z')$ that are intrinsically twice as sensitive to metallicity as $(V-I)$.
All of these factors bring many more galaxies within reach for GCS
studies and allow us to define their systemic properties at a completely
new level of confidence.

Throughout this paper, we use a distance scale $H_0 = 70$ km s$^{-1}$
Mpc$^{-1}$ to convert redshifts and angular diameters to true distances.

\section{Observations and Data Reduction }

In this paper we present the color distributions and MDFs for eight
BCGs that we have observed as part of HST program GO 9427.  The targets are
listed in Table 1, presented in approximate order of increasing distance.  
All of them are giant ellipticals at the centers
of moderately nearby clusters, with redshifts ranging from $\sim 1800$
to 3200 km s$^{-1}$.  Some (NGC 1407, 5322, 7049) are in relatively sparse
clusters, others (NGC 3258, 3268, 4696) in quite rich ones.  Some have
evidence for recent merger or accretion activity, such as dust lanes or
other complex, small-scale features in the innermost $\sim 1$ kpc
around the nucleus, but in all cases the halo and bulge regions well
beyond the nucleus are clear and uncrowded, permitting easy measurement
of the globular cluster populations.

Table 1 lists in successive columns (1) the target NGC number, (2) the galaxy
group or cluster in which it is the centrally dominant gE (NGC 3258 and 3268 
are a rare example of a co-dominant central pair), (3) the redshift $cz$ of
the galaxy, corrected where possible to the CMB reference frame
\citep{ton01}, (4) the galaxy luminosity $M_V^T$ as listed in the NASA
Extragalactic Database (NED), (5) the foreground
reddening, from \citet{sch98}, (6) the apparent distance modulus in $I$,
calculated from $cz$, $H_0 = 70$, and the adopted reddening, and
(7,8) the exposure times
in the $F435W$ and $F814W$ filters, in seconds.
For NGC 4696 in particular, the foreground reddening is 
somewhat uncertainly known:  following \citet{bur03} and \citet{blake01}, we adopt
$E_{B-I} = 0.24$ instead of the Schlegel et al. value of 0.28.  For its
distance, we also use the more recent value
$(m-M)_0 = 33.11$ from \citet{mieske03}, slightly lower
than the Tonry et al. compilation.

For each target we obtained exposures with the ACS Wide Field Camera
in filters $F435W$ and $F814W$ (broadband $B$ and $I$). Each galaxy target
was centered either near the middle of one of the ACS/WFC CCDs, or near
the geometric center of the whole array.  The exposure
times were designed to
be deep enough to reach past the turnover (peak frequency) of the
globular cluster luminosity function (GCLF) at $M_{V,to} \simeq -7.4$ 
\citep{h01} at high detection completeness.  
This luminosity is equivalent to $M_{I,to} \simeq -8.4$ for a mean cluster
color of $(V-I)_0 \simeq 1.0$.  

In the present paper, we present the two-color
data and some intriguing new findings about the metallicity 
distributions.

\subsection{Photometry of the ACS/WFC Images}

For the data reductions we started with the Multidrizzled frames
provided by the HST pipeline preprocessing and extracted from the
HST Archive.  These images are corrected for the geometric distortion
of the camera, reoriented to the cardinal directions on the sky, and
are cleaned of cosmic rays as far as allowed by the small number
of exposures in each orbit as listed in Table 1.  These images have a 
renormalized scale of $0\farcs05$ per pixel.  In all cases the $B$ and $I$ images were
well registered with each other, requiring only simple $xy$ shifts
of 1 pixel or less to match up their internal coordinates.

\begin{figure}
\plotone{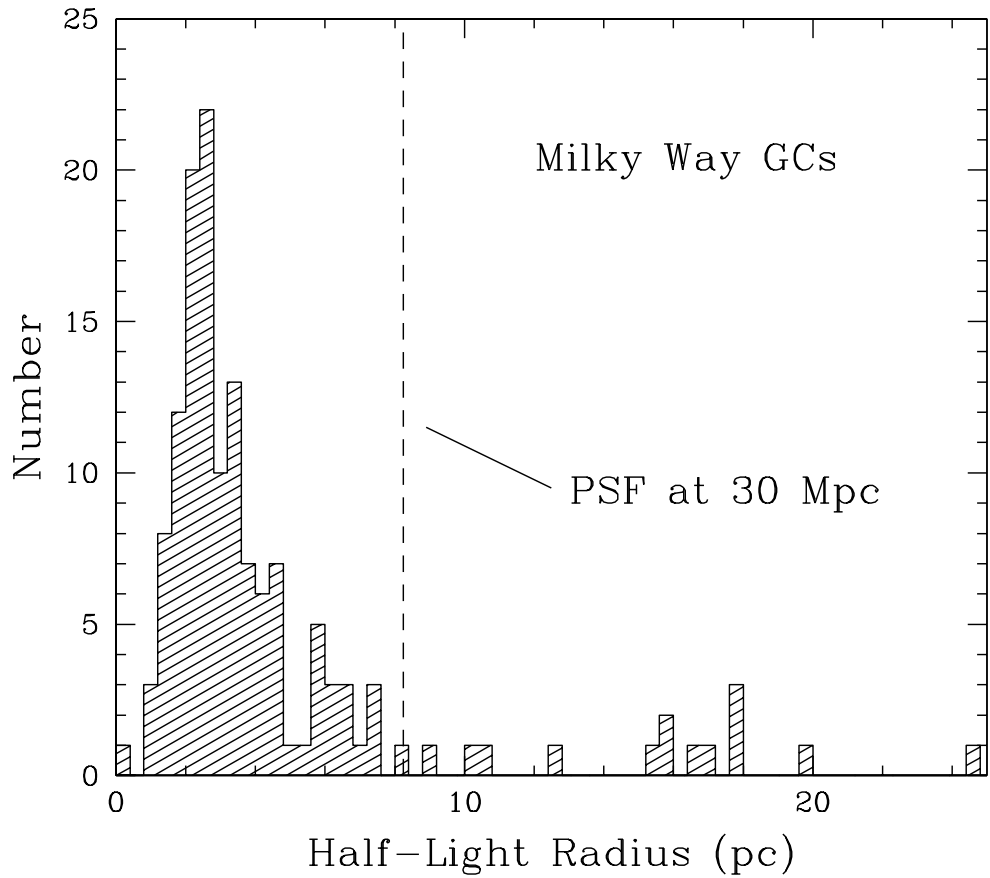}
\caption{Distribution of observed half-light radii $r_h$ (in parsecs)
for the Milky Way globular clusters, with data from Harris (1996).
For the ACS images used in this study, the stellar point spread function
(PSF) has a typical HWHM of 1.1 pixels, which is equivalent to 8.3 pc
for a galaxy at a distance of 30 Mpc, as shown by the dashed line.
}
\label{halflight}
\end{figure}

For the photometry we used the standalone DAOPHOT suite of codes
\citep{stet94} in its most recent version {\sl daophot-4}.  
The reductions followed the normal {\sl find/phot/psf/allstar} sequence to
find objects above the adopted detection threshold (3.5 to 4.0 times the
standard deviation of the sky background), carry out small-aperture
photometry, define the point spread function, and fit the PSF to all
the objects in the detection list.  For galaxies at these distances,
the globular clusters we are trying to find
are nearly starlike:  a typical
GC half-light diameter $r_h \simeq 5$ pc (Figure \ref{halflight})   
will subtend $0\farcs035$
(less than 1 pixel) at the $\sim 30$ Mpc distance typical of our
target galaxies, whereas the stellar FWHM on the multidrizzled
images is $2.0 - 2.5$ pixels.  An average GC profile convolved with the
stellar PSF would then increase the FWHM by 10\% or less.
Thus we have found it readily possible for our present
purposes to carry out PSF-based photometry on our images.  This approach
works particularly since
the majority of bright, ``starlike'' objects in the frames are actually
globular clusters scattered around their galaxies (see below), and we use
these to define the mean PSF for each frame.  Thus, to a large extent
we use the clusters themselves to define the mean PSF profile.
In the later discussion we will, however, describe an additional
check on the PSF-based photometry through aperture photometry.

Most of the objects on our 
images are uncrowded (the  typical nearest-neighbor separation
even near the galaxy centers where the clusters are most 
numerous) is $\sim 2'' - 3''$ (40 -- 60 pixels), yielding a very
high detection completeness through a single pass of {\sl daophot/find}
with appropriate choices of detection threshold.

\begin{figure}
\plotone{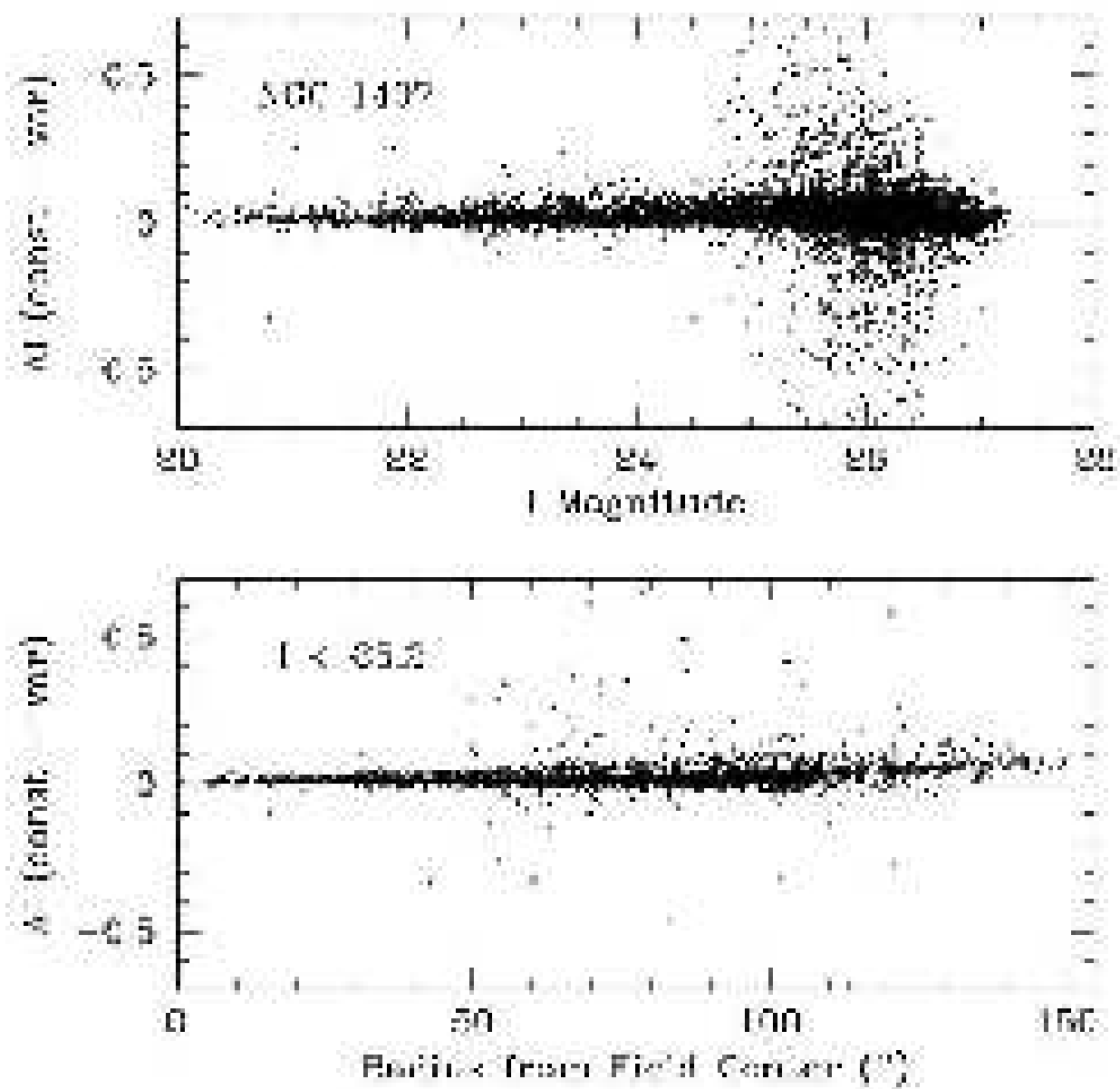}
\caption{({\it Upper panel:}) Comparison of photometry
for a constant PSF versus a PSF with quadratic dependence
on field location $x,y$.  For the NGC 1407 $I-$band image, the difference between
  the two measurements is plotted against $I$ magnitude.
The enhanced scatter near $I \sim 26$ is due mainly
to very faint objects near the galaxy center where the background
light is high.
  ({\it Lower panel:}) Difference plotted versus radius from the
center of the ACS/WFC field, in arcseconds.  Most of the stars follow
a trend of gradually increasing $\Delta I$ with radius; a more thinly
populated ``second sequence'' sitting at higher $\Delta I$ belongs to
stars located along one particular radial direction (detector corner).}
\label{psftest}
\end{figure}

We defined the point spread function empirically for each frame,
using anywhere from 50 to 80 
``stars'' distributed across the frame (as noted above, many of these are
bright globular clusters, with measured FWHMs of $2.1-2.5$ px, a range
similar to that generated by the field dependence of the PSF).
The candidates were manually selected to be
bright and uncrowded.  Initially
we ran trials for (a) a constant PSF (i.e. the same everywhere on
the frame), (b) a PSF linearly variable in $x$ and $y$, and
(c) a PSF quadratically variable in $x$ and $y$.  The
quality of fit across the images was marginally but noticeably 
best with the quadratically variable PSF, though it produced
very little difference in the measured magnitudes of the objects
anywhere in the field.  A sample of these tests is shown in Figure 
\ref{psftest}.  No trend with magnitude between the two types of PSFs
is apparent, and only a slight trend with field location, 
reaching a center-to-edge amplitude $\simeq 0.05$
mag at the corners of the frames.  (The very slight zeropoint offset 
between the two PSF scales is not significant, since it is routinely
removed during the aperture-correction stage).

For each galaxy, we constructed a master finding list of objects by
reregistering and stacking all the available multidrizzled frames in 
both colors, and running {\sl daophot} on the summed frame.  Then, we carried
out {\sl allstar} with the master candidate list
on each of the $B$ and $I$ frames, keeping only
objects that appeared on all frames; this step rejected virtually
all remaining cosmic rays and other artifacts that might have crept 
through the Multidrizzle preprocessing, as well as any objects with
extreme colors.

Lastly, we have avoided any objects located within $5''$ of the galaxy
nuclei, because of the bright background light there.  This region corresponds
typically to a linear radius less than 1 kpc for our target galaxies, 
and has little effect on any of the following discussion.

\subsection{Calibration Steps}

Following the prescriptions in \citet{sir05}, we used bright starlike
objects (essentially, the bright clusters used to define the PSF) on each frame
to determine the mean ``aperture correction'', which is the difference
$(m_{psf} - m_{10})$ between the psf-fitted magnitude and the magnitude
measured through a 10-pixel-radius ($0\farcs5$) aperture.  We then 
extrapolated from $0\farcs5$ to ``infinite'' radius subtracting 
the values recommended by Sirianni et al. 
($\Delta B = 0.107, \Delta I = 0.087$ mag).  In summary, the
measured magnitudes on the natural VEGAMAG filter systems are then
\begin{eqnarray}
F435W = b_{psf} + \Delta B_{ap} - 0.107 \\
F814W = i_{psf} + \Delta I_{ap} - 0.087  
\end{eqnarray}
\noindent where $b,i_{psf}$ are the magnitudes measured by {\sl daophot/allstar},
normalized to an exposure time
of 1 second, and $\Delta B,I$ are the corrections to an aperture magnitude
of radius 10 pixels.

The last step is to convert the instrumental magnitudes to standard $B$ and $I$.
We have adopted the synthetic model prescriptions from \citet{sir05} for objects
in the color range $(B-I) > 1$, which we repeat here:
\begin{eqnarray}
F435W = B - 25.749 - 0.008 ~ (B-I) + 0.005 ~ (B-I)^2 \\
F814W = I - 25.495 + 0.010 ~ (B-I) - 0.006 ~ (B-I)^2 
\end{eqnarray}
\noindent Iteration of these relations quickly converges to a final pair of $B,I$
magnitudes, since the color terms in the transformations are very small.

We have not applied any CTE (charge transfer efficiency) corrections to the
raw photometry.  Because these images are relatively long exposures with 
broadband filters, and are located in the outskirts of bright giant galaxies, the
ambient sky background is high enough that CTE corrections are 
expected to be 0.02 mag or less for all our data \citep{riess04}.

Combining all the calibration steps,
we believe our $B$ and $I$ zeropoints are likely to be uncertain
from one galaxy to another by $\pm 0.02 - 0.03$ mag each.  The dominant
sources of error are the aperture
corrections ($\Delta B, \Delta I$), the extrapolations to infinite
aperture radius, and the zeropoints in the final transformation
equations, each of which appear to have potential systematic
uncertainties of $\pm 0.01 - 0.02$ mag.  In addition to these
calibration uncertainties, the {\it external}
uncertainties in the foreground reddenings of the galaxies in
our survey are at the level of $\Delta (B-I) \sim \pm 0.02$ mag.  These small potential
shifts are what place ultimate limits on the degree to which we
can intercompare the $(B-I)$ distributions even within one strictly homogeneous
dataset.

\subsection{Aperture vs. PSF Photometry}

To test the robustness of the color-magnitude data to
the details of the photometric procedures, we have extensively
compared the {\sl allstar}
PSF-fitted data with magnitudes measured directly from small-aperture
photometry. 
The main advantage of aperture photometry is that it provides a simple measure of
the total light within a fixed radius and is less affected by
small object-to-object differences in the actual profile shape. The
compensating advantage of PSF fitting is that it is less affected by crowding.
For the very faintest objects, PSF fitting through {\sl allstar} is also less
subject to the occasional ``wild'' excursions 
in measured magnitude that can affect aperture photometry due to combinations 
of noise in both the star and sky annuli, crowding, and centering
uncertainties.
For this comparison, we used a 3-pixel aperture radius (similar to the
PSF fitting radius) and calibrated it
to the absolute $B$ and $I$ magnitude scales with the same techniques 
described above.

In Figures 3 to 6, we show the color-magnitude plots  in $(I, B-I)$ for all eight of our
BCGs, with both PSF-based {\sl allstar} photometry and with small-aperture
photometry.  In each pair of graphs the PSF photometry is shown on the left, and
the aperture photometry of exactly the same set of objects on the right.

The main point we draw from this comparison is that the
two methods give extremely similar color-magnitude distributions, essentially
because of the fundamental advantage that 
crowding is not a major problem in any of these fields.
In each galaxy field, the brighter part of the color-magnitude array
is populated by a pair of near-vertical sequences in the color
range $1.4 < (B-I) < 2.6$ that are the 
red and blue globular cluster populations.  The detailed characteristics
of these sequences will be discussed in later sections.
In all the diagrams the main
contribution from field contamination comes in at faint, blue levels
($I > 25, (B-I) < 1.5$); although contamination from 
similarly faint red objects must exist, these objects are mostly cut off by
the faint-end incompleteness in the $B$ images and thus do not appear in
the diagrams.

We find no systematic differences in magnitude or color
between the two methods larger than the $\pm 0.02-$mag calibration uncertainties
discussed above.  As expected,
the aperture-based photometry tends to show (a) a few more ``outliers'' at very red
or blue colors produced by occasional crowding; and 
(b) a very slightly larger color spread in the red and blue cluster sequences.  
For these reasons, we prefer to use the PSF-based photometry in the rest of
our discussion.

The single exception to this choice is the NGC 4696 field, which is the one with
the largest measured number of clusters in our survey.
Close inspection of Figure 6 shows 
a second-order effect on the mean colors for objects fainter than $I \sim 23.5$.
Within the PSF-based photometry, the average color of the cluster population
becomes slightly and progressively bluer toward fainter magnitudes, whereas
in the aperture photometry the mean colors of the cluster sequences remain
constant with magnitude over that $I$ range.
We have not been able to find a definitive explanation for this small but 
puzzling trend in the PSF photometry for this one field, but 
it may be due either to (a) higher-order nonuniformities
in the PSF shape across the frame that we have not been able to track, or
(b) small differences among the true radii 
of the clusters (thus their true profile widths).
Fully accounting for this residual
effect will require much more extensive and painstaking object-by-object photometry. 
For the present, we have chosen to use the aperture-based photometry for NGC 4696
in the discussion that follows.  For the other galaxies, we use the PSF data.
As will be seen below, however, much of our discussion is based on the
{\sl brighter} clusters in the color-magnitude data and thus is little affected
by this particular issue.

\subsection{Conversion of Color to Metallicity}

Although the $(B-I)$ index is one of the most effective ones for GCS metallicity
estimation that can be constructed from commonly used broadband filters, it has
only rarely been employed before in GCS studies \citep{har90,grillmair99,whit03,jordan04}.
A necessary preparatory step for our analysis is to define a conversion
from color to metallicity,
and although the globular clusters in the Milky Way are not guaranteed to
be identical to those in giant ellipticals, they remain
our best resource for defining the conversion.
In Figure \ref{bifeh}, we show the catalog data \citep{har96} for 95
Milky Way clusters with known $(B-I)$, spectroscopically 
measured [Fe/H], and foreground reddenings
less than $E_{B-V} = 1.0$.  We adopt a linear conversion
\begin{equation}
(B-I)_0 = (2.158 \pm 0.068) + (0.375 \pm 0.049) ~ [Fe/H]
\end{equation}
shown by the solid line in the Figure.  The uncertainty in the slope
is such that for the very reddest or bluest clusters,
the deduced [Fe/H] could be systematically in error by $\pm 0.1$ dex.
In addition, any cluster metallicities higher than Solar require
extrapolation of the relation and thus are only estimates.

\section{The Color-Magnitude Data }

\begin{figure}
\plotone{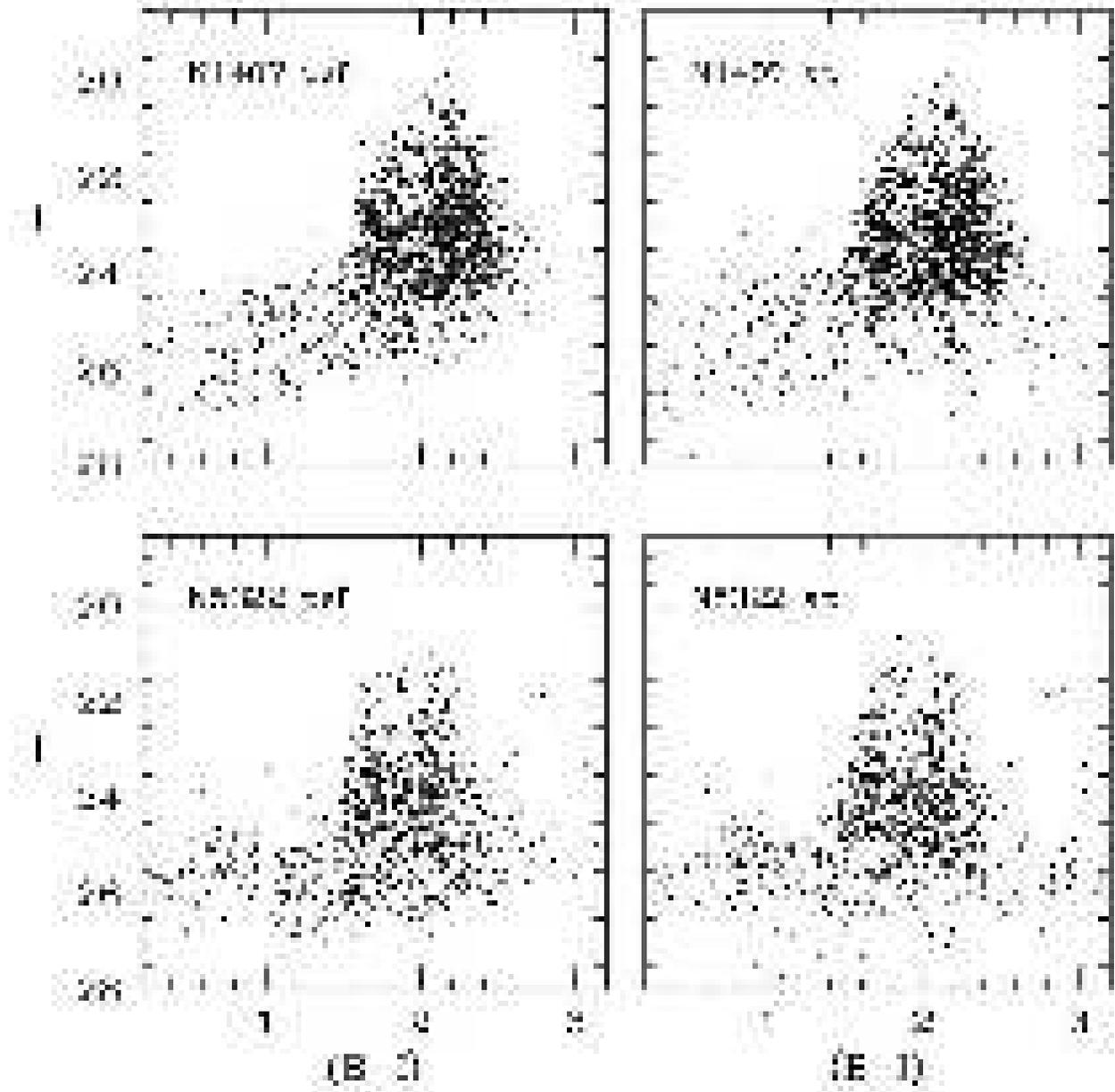}
\caption{Color-magnitude plots for the measured objects around NGC 1407
(top) and NGC 5322 (bottom).  In each pair of plots the {\sl left panel}
shows the photometry measured via {\sl allstar} PSF profile fitting,
while the {\sl right panel} shows the same objects measured through
fixed-aperture photometry of 3 pixels radius.  No corrections for reddening
have been applied to the data.
The objects in the approximate color range $1.4 < (B-I) < 2.6$ are in the
expected range for old globular clusters.}
\label{cmd4a}
\end{figure}

\begin{figure}
\plotone{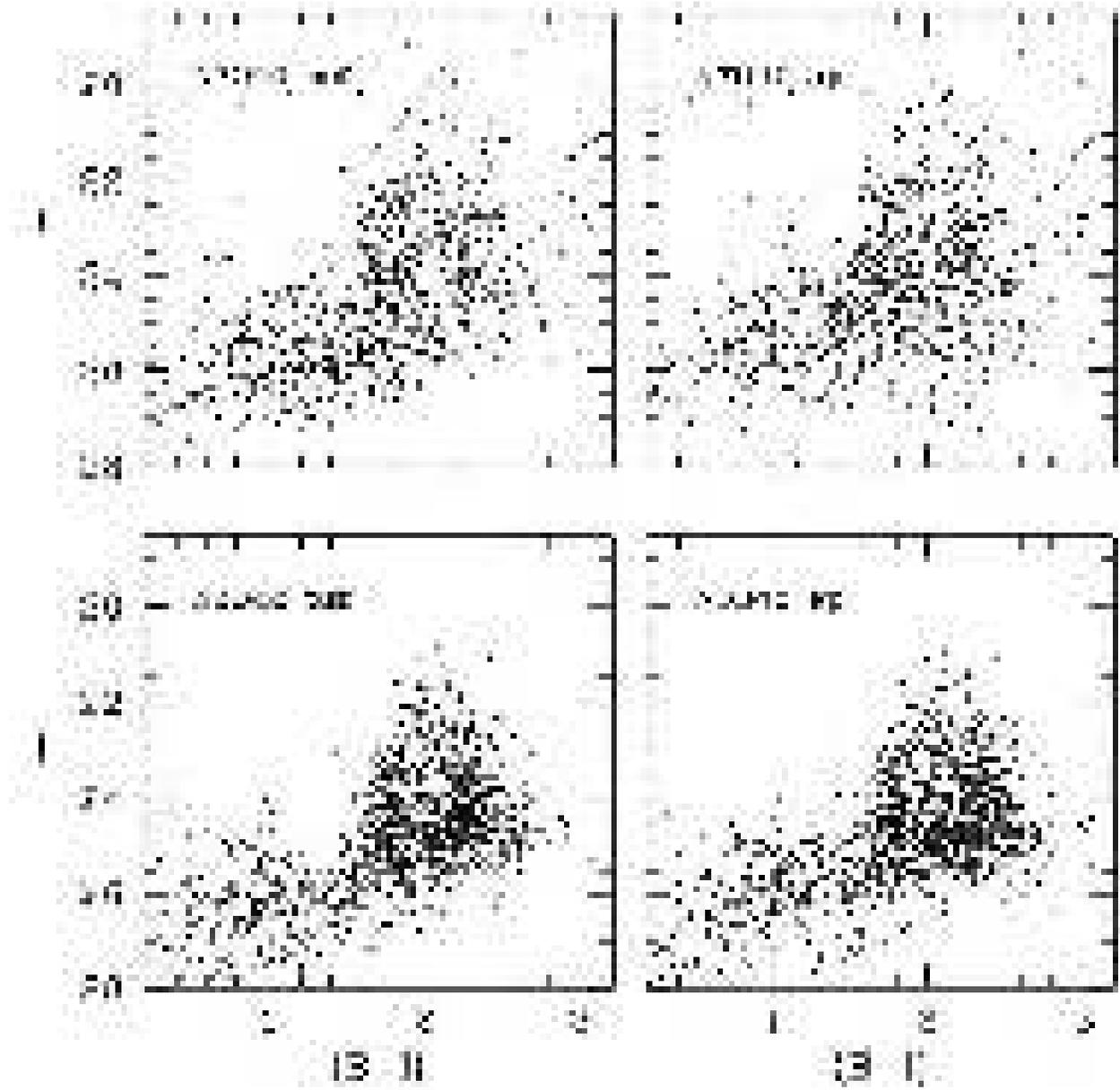}
\caption{Color-magnitude plots for the cluster populations around
NGC 7049 (top) and NGC 3348 (bottom).}
\label{cmd4b}
\end{figure}

\begin{figure}
\plotone{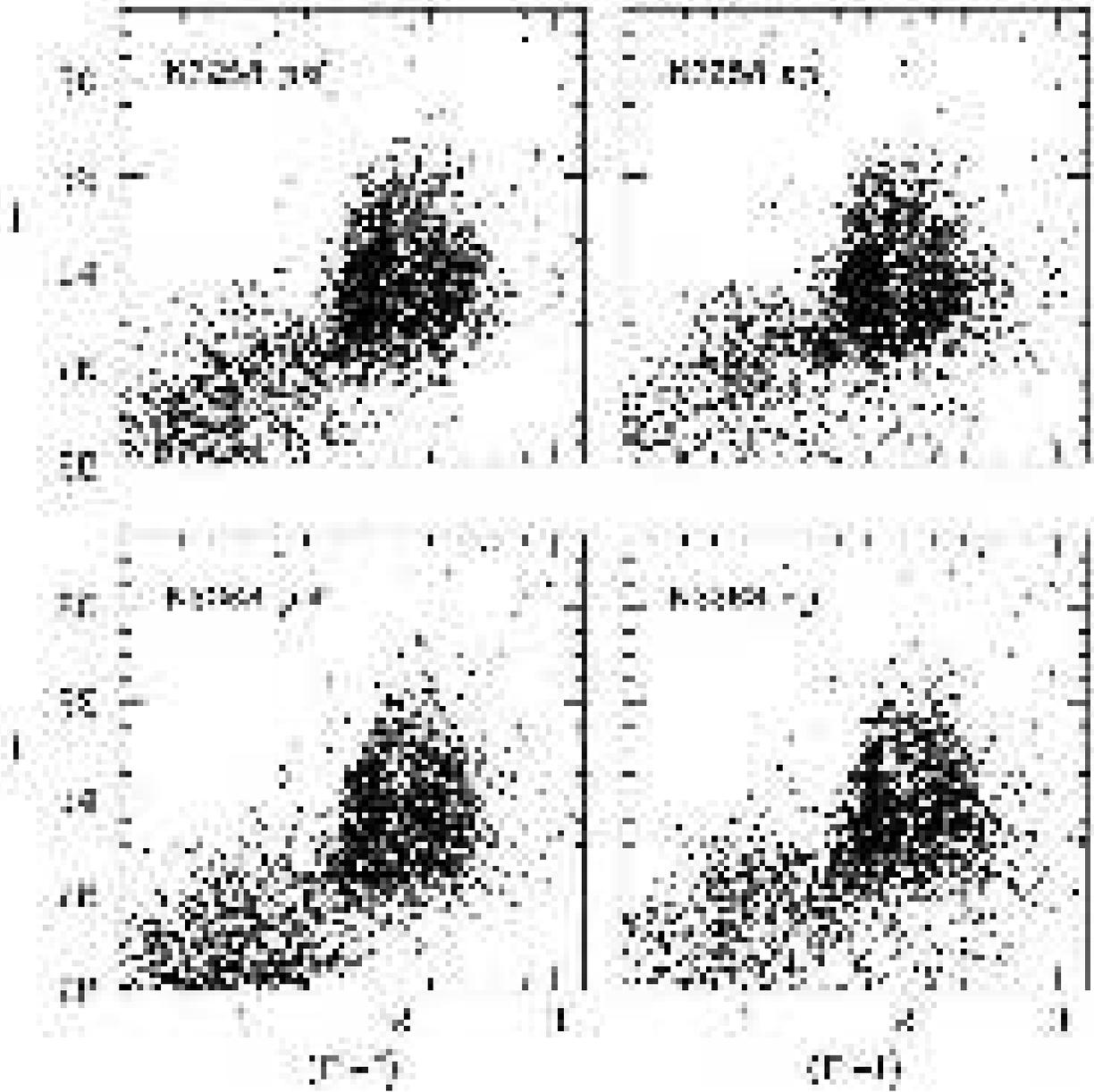}
\caption{Color-magnitude plots for the cluster populations around
NGC 3258 (top) and NGC 3268 (bottom).}
\label{cmd4c}
\end{figure}

\begin{figure}
\plotone{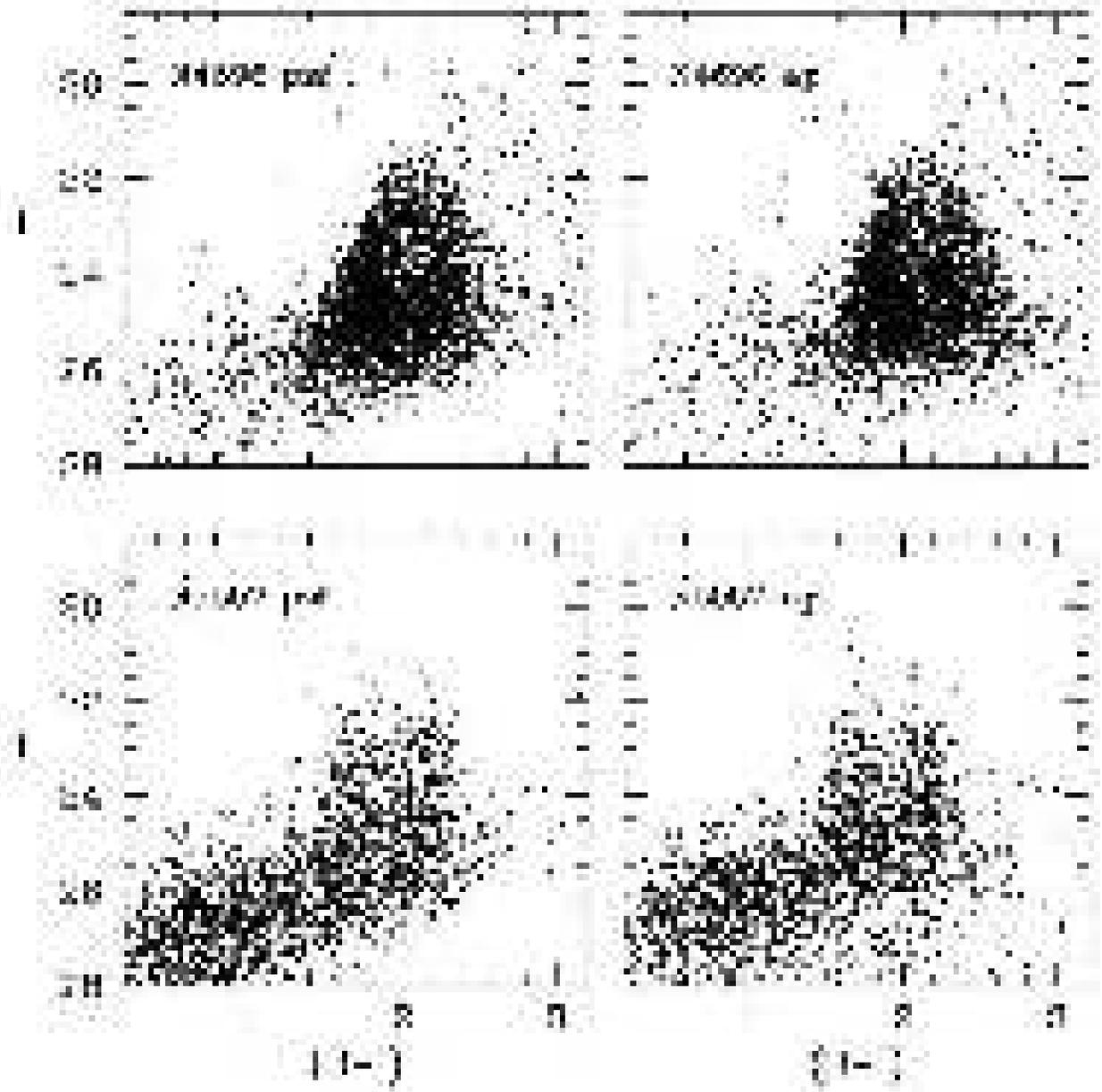}
\caption{Color-magnitude plots for the cluster populations around
NGC 4696 (top) and NGC 5557 (bottom).}
\label{cmd4d}
\end{figure}

\begin{figure}
\plotone{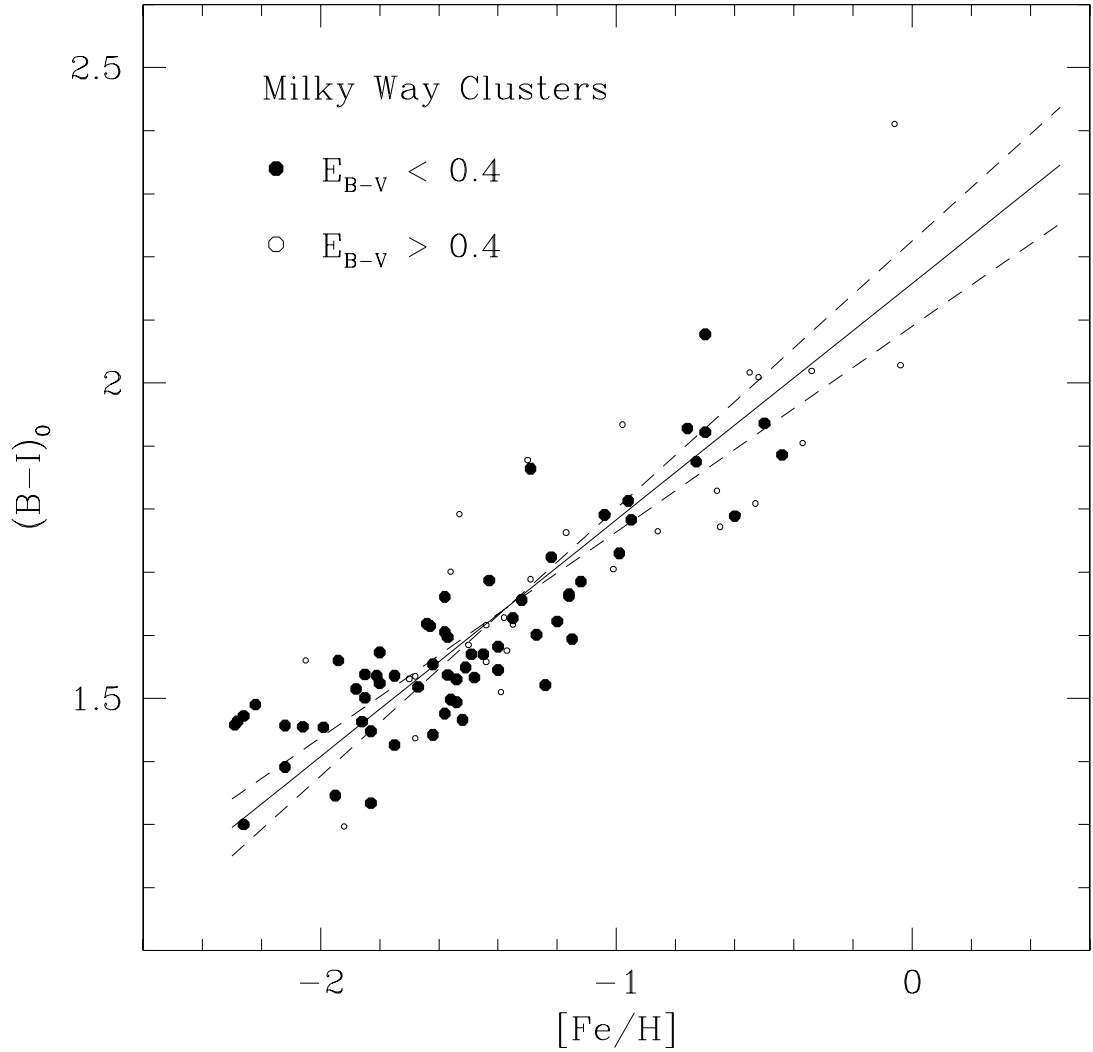}
\caption{Intrinsic color $(B-I)_0$ versus [Fe/H] for
Milky Way globular clusters, taken from the 2003 edition of
the Harris (1996) catalog.  Clusters with lower reddenings
($E_{B-V} < 0.4$) are plotted as large solid symbols, and
ones with $0.4 < E_{B-V} < 1.0$ as small open symbols.
The two dashed lines show the two linear least-squares solutions
([Fe/H] vs. $(B-I)_0$, or $(B-I)_0$ vs. [Fe/H]); the solid line
is the adopted conversion as given in the text.}
\label{bifeh}
\end{figure}

In Figures \ref{cmd4a} to \ref{cmd4d}, the $I$ vs. $(B-I)$ 
scatter plots (color-magnitude diagrams or CMDs) for
all eight BCGs are shown.  The first and most obvious features to emerge from 
these plots are that (a) the GCS populations stand out clearly in
all the galaxies as the vertical sequences within $1.4 < (B-I) < 2.6$,
(b) the color distributions are {\sl strongly and unequivocally 
bimodal} in all the galaxies, and (c) the relative numbers of blue and
red clusters differ noticeably from one galaxy to the next.
We will
quantify these characteristics in the following discussion, but at
this point it is simply worth noting that the promise held out by
using the combination of ACS/WFC with long exposure times and with
the $(B,I)$ filters has
been fulfilled, removing all the problems associated with
the previous, more limited $(V-I)$ studies.  
The blue and red sequences are more clearly separated
in these new data than in all but the best photometry in the previous literature,
and as will be shown below, the internal dispersions in each sequence
are well resolved.

\subsection{Estimates of Field Contamination}

Field contamination of the CMDs -- that is, the residual presence
of foreground field stars and very faint, compact background galaxies that 
succeeded in passing through the photometric procedures -- is clearly at
a low level, at least for the brighter
parts where the cluster sequences stand out the most
strongly.  At fainter levels (corresponding to $M_I > -9$ in rough terms),
large numbers of faint blue objects populate the diagrams; similarly faint but red
objects are visible in the $I$ images, but they are
cut off by the detection incompleteness in the $B$ frames and thus
do not appear in our two-color data.  However, the data at these fainter levels are of no
interest in our present discussion and are not used here in any way.
They are more relevant to the derivation of the globular cluster luminosity
functions (GCLF) which will be the subject of a future paper.

The magnitude and color ranges of interest for our present work are the ones
which bracket the brighter globular clusters, $-13 < M_I < -9$ and
$1.4 < (B-I)_0 < 2.4$ (see especially Figure 11 later in the paper for
CMDs plotted in $(M_I, (B-I)_0)$).  This is the region
for which we are more directly interested in the level of contamination.
To quantify the true number of contaminating objects very 
exactly, we would need to have remote-background fields to refer to, but these
are not on hand.  Color-magnitude
plots of each field subdivided by galactocentric radius also show that the globular cluster
sequences are prominent even out to the edges of the ACS camera field.  That is,
the GCSs in these supergiant galaxies extend detectably well past the borders of
our fields, so that we cannot use the outer parts of the frames to set anything
but the most generous (and in practice, useless) upper limits on the field
contamination.

What we have done instead to {\sl estimate} the level of contamination is to
refer back to the color-magnitude diagrams themselves.  We use the numbers
of stars that appear in the same luminosity range ($-13 < M_I < -9$)
as the brighter globular clusters, but
at {\sl bluer or redder colors} than the one-magnitude 
range ($1.4 < (B-I)_0 < 2.4$) that encloses the globular clusters.  If we 
then assume more or less arbitrarily that field stars are evenly spread in
color over the wider range $\simeq 0.8 - 3.2$ in $(B-I)_0$, then the counted
number of stars per unit magnitude will give a plausible estimate of the
contamination level within the globular cluster sequences.

The results of this numerical exercise are shown in Table \ref{field}.
The successive columns give (1) NGC number, (2) total number of globular cluster
candidates in the range $-13 < M_I < -9$ and $1.4 < (B-I)_0 < 2.4$; 
(3) estimated number of contaminants in the same range, from the number of bluer
and redder objects present in the CMD; and (4) the ratio $N_c / N_f$.
It is clear from these numbers that the contamination level is only a few
percent for most of the fields, reaching a maximum of 11\% for the sparse
NGC 7049 GCS.  We conclude that we can safely analyze the color distributions
of the candidate GCS populations without field contamination corrections.

\subsection{Bimodality Model Fits}

The raw color-magnitude diagrams for the eight BCGs can now be used
to determine more quantitative features of the color/metallicity
distributions of the globular clusters.
The first and most basic issue is to determine how strongly a bimodal
fit is preferred at any level in the CMDs over simpler models such as
a unimodal distribution.  This question is easily answered for most
of the visible luminosity range, but less so at the faintest levels
in our data where the red and blue sequences become lost in 
increasing measurement errors and field contamination, and also at the
top end (discussed especially in the following section).

Unlike the custom in many previous studies,
we do not assume equal dispersions in these KMM fits for the
blue and red subpopulations; such an imposition on the data is 
unjustified when the data are of high enough quality to resolve
differences between the modes. 

\begin{figure}
\plotone{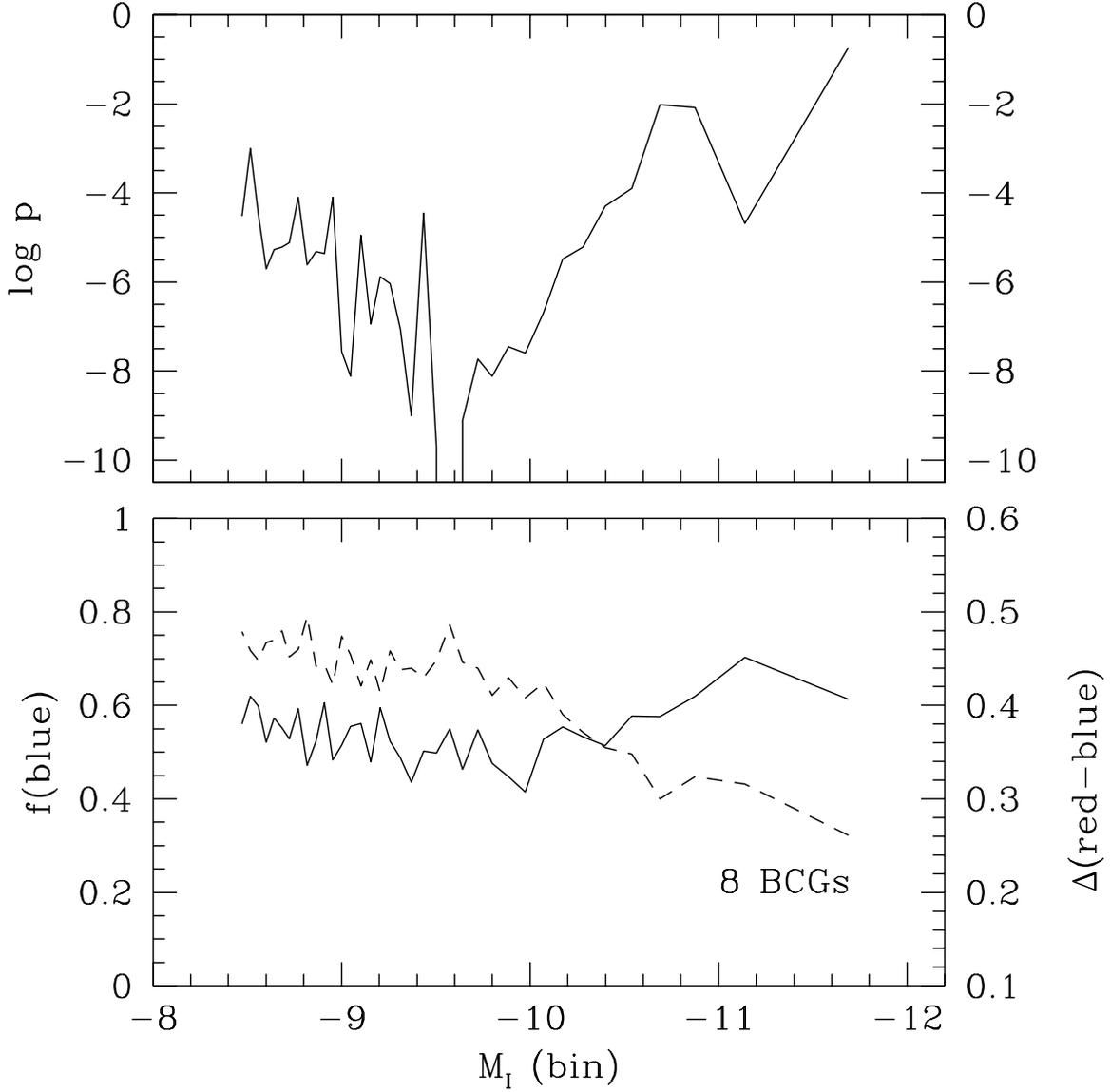}
\caption{{\sl Upper panel:}  Results of the KMM 
model fits to the combined color distribution for 
all eight BCGs, subdivided into narrow luminosity bins.  Here,  $p$ is the probability
that the color distribution within each bin
can be represented by a single (unimodal) Gaussian; the logarithm of $p$ 
is plotted against the mean luminosity of the objects in the bin, $\langle M_I \rangle$.  The luminosity
intervals are defined so that 200 objects are in each bin.
{\sl Lower panel:}  Solid line shows the fraction of clusters in the blue 
mode, plotted as a function of luminosity bin (ordinate on left side).  
The dashed line shows the
difference $\langle B-I\rangle$(red) $- \langle B-I\rangle$(blue)
between the peak colors of the two modes (ordinate on right side).
}
\label{all8_200}
\end{figure}

In Figure \ref{all8_200}, we show the results of this exercise where
we have combined all eight BCGs into a single color-magnitude distribution
in $(M_I, (B-I)_0)$, and divided the combined list into narrow luminosity
intervals such that precisely 200 objects are in each bin.  The upper
panel of the Figure shows the probability $p$ given by the KMM fitting
code that the color distribution can be matched by a single Gaussian curve.
For all luminosities $M_I > -10.5$, we find log $p$ $< -3$ and the
null hypothesis of unimodality can therefore be strongly rejected.  Two other
features of this graph, however, are worth noting: first, for $M_I > -9.5$
the value of $p$ slowly increases towards fainter levels from 
$\sim 10^{-10}$ to $10^{-4}$. This trend shows
the increasing effects of photometric 
measurement scatter and some field contamination that tend to make the
blue and red modes spread out and partially overlap.  Second, for
the upper range $M_I < -10.5$, a few of the bins have 
probability levels $p > 0.01$, 
which means that a unimodal-Gaussian fit to the color distribution
starts to become a plausible (though still not convincing) fit.
This effect cannot be attributed to either photometric scatter or
field contamination, and must be regarded as real.
This particularly interesting top end of the CMDs will be discussed
in more detail in the next section.

The lower panel of Fig.~\ref{all8_200} shows (solid line) the trend of
$f$(blue), the fraction of clusters that the KMM fitting routine
assigns to the blue mode (and where by definition $f$(red) $\equiv 1 - f$(blue)). 
If the clusters in both modes have intrinsically similar mass distributions,
then we would expect to see little systematic change in this ratio
with luminosity.  Indeed, little if any significant change in the ratio
takes place, with on average about 55\% of the whole GC sample falling
into the blue, metal-poor mode.  Lastly, in the same figure, the dashed line shows the 
color {\sl difference} $\Delta \langle B-I\rangle$  between the peaks of two fitted Gaussian
curves.  This difference shows no change with luminosity for $M_I > -9.5$, but
above that it steadily decreases toward higher luminosity, indicating that
the two fitted modes are gradually converging toward the top end (see below).

\subsection{Mean Colors and Dispersions}

\begin{figure}
\plotone{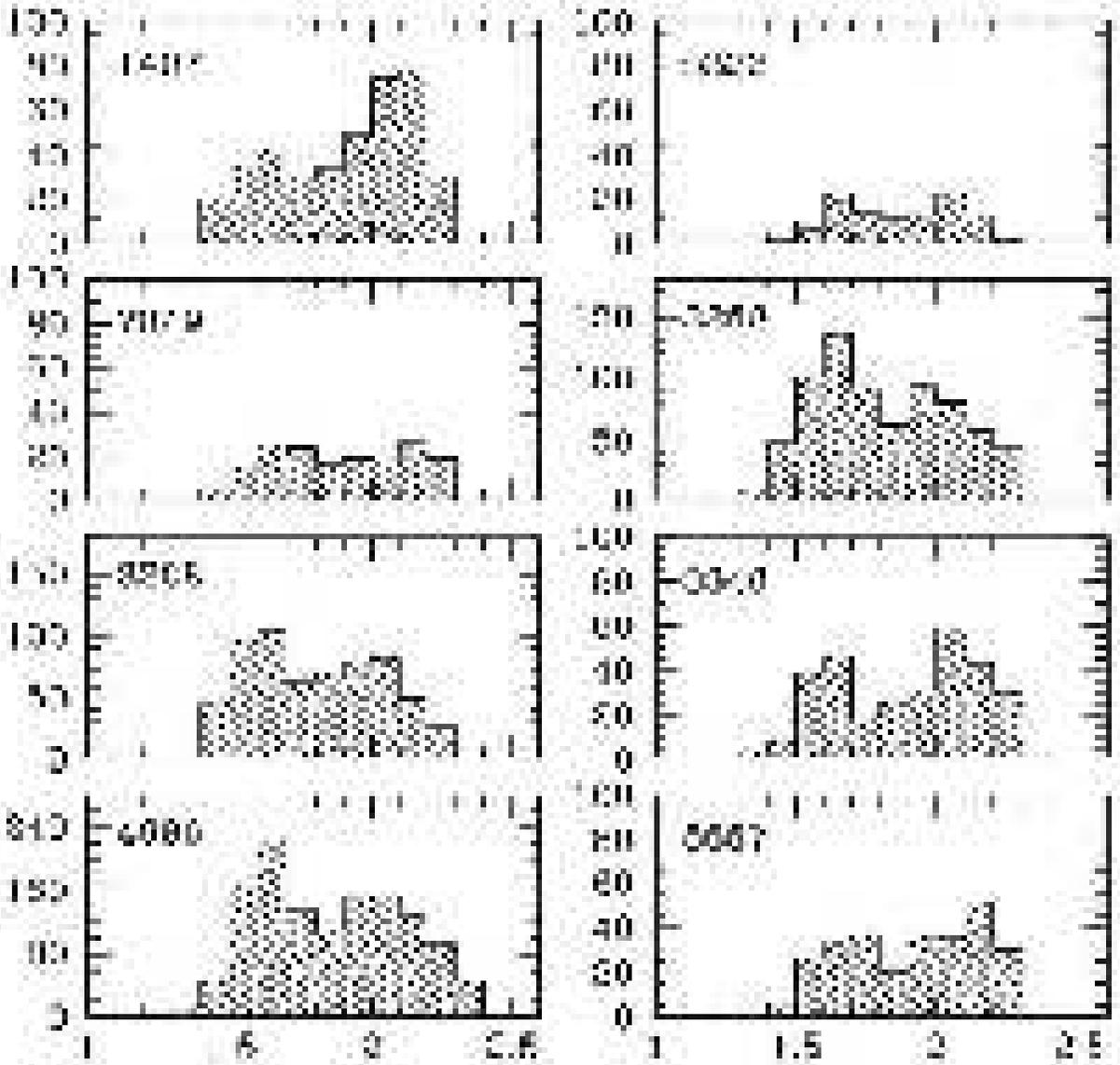}
\caption{Histograms in color for the globular cluster systems.  Here, the number
of objects per 0.1-mag bin in $(B-I)_0$ in the luminosity range 
$-10.5 < M_I < -9.0$ is shown for each galaxy.  All the distributions show
the basic phenomenon of bimodality.
}
\label{bihisto8}
\end{figure}

The dereddened color histograms derived directly from the color-magnitude plots 
are shown for each galaxy in Figure \ref{bihisto8}.  Here,
we use only the objects in the specific luminosity range
$-10.5 < M_I < -9.0$; as we note above, the faint end is set to minimize effects of field
contamination as well as the spreading effect of photometric random errors. 
The bright end is set for reasons that were hinted at above, indicating that
this upper end requires special treatment.
We have used these histograms to determine
parameters for best-fit double Gaussian curves derived by both a 
maximum-likelihood method through the 
KMM mixture modelling routine \citep{ash94}, and independently by a separate
nonlinear least-squares Levenberg-Marquardt code.  The results from
these two independent codes agreed closely, and we will refer to only
the KMM fits below.

One of the first points to emerge from our 
new MDFs is that the red population consistently displays
a significantly bigger color dispersion (thus metallicity range) than the blue population.
In Table 3, we list the parameters of the resulting fits, as well as the
fitted ratio of total subpopulations $(N_{blue}/N_{red})$, and the color
difference $\Delta \langle B-I \rangle$ between the red and blue
modes as defined above.  The mean values over the eight galaxies are listed at the bottom
of the Table, along with the galaxy-to-galaxy rms dispersions (in parentheses).

\begin{figure}
\plotone{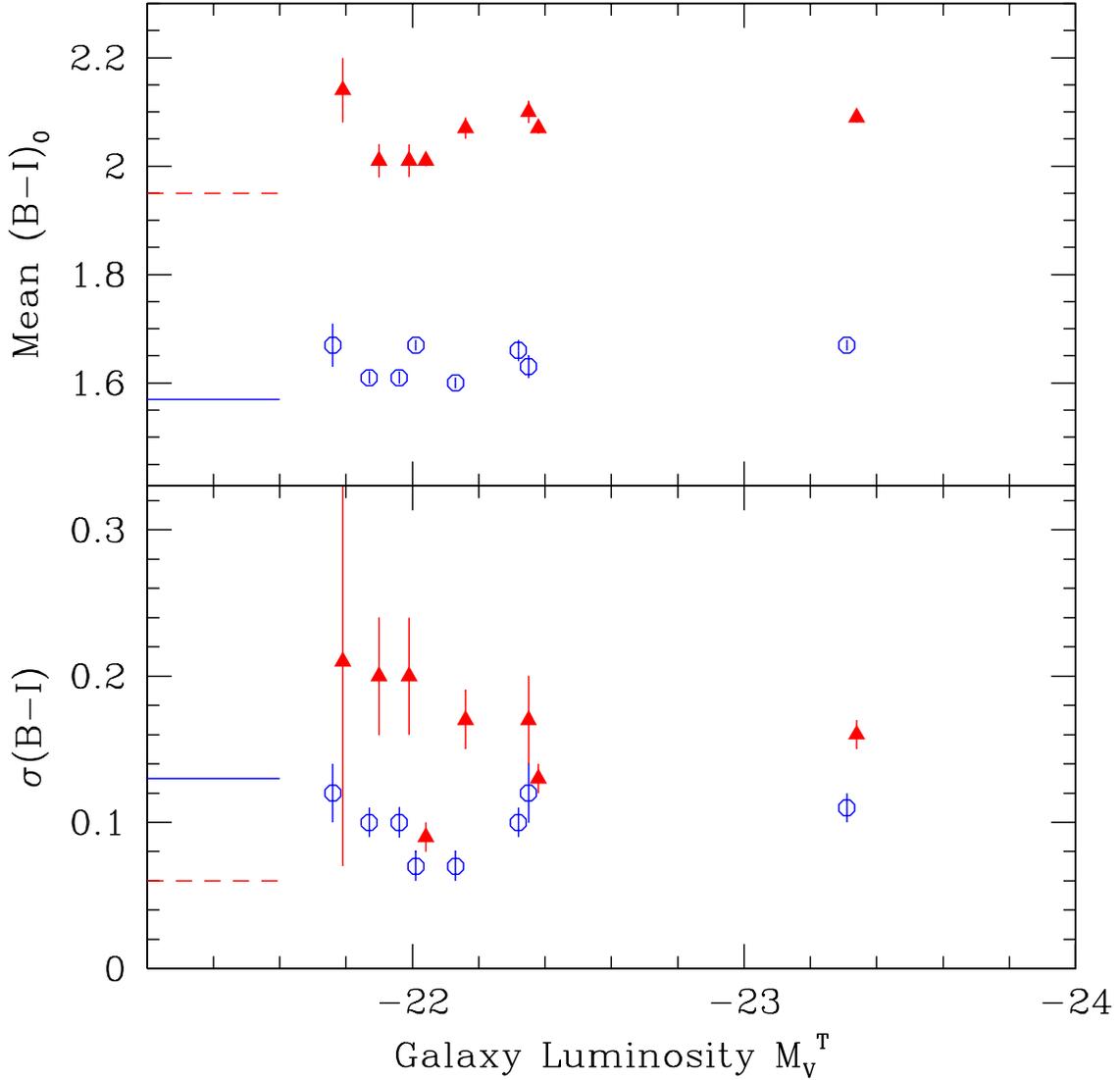}
\caption{Parameters for the double-Gaussian fits to the color histograms in
Figure 8, with data from Table 3.  
The Gaussian peaks $\langle B-I \rangle_0$ and standard
deviations $\sigma_{B-I}$ of the red and blue modes
are plotted versus galaxy luminosity.  Points for
the blue (low-metallicity) clusters are the open circles and the red
(high-metallicity) clusters are the filled triangles.  
The error bars in the upper panel show only
the internal uncertainty of the mean and do not include the uncertainty in the foreground
reddening of the galaxy itself.  The mean and standard deviation
for the metal-poor Milky Way globular clusters are indicated by the 
{\sl solid lines}, while the metal-rich Milky Way clusters are shown by the
{\sl dashed lines}.  Note that the metal-rich Milky Way clusters form a
considerably narrower subpopulation (smaller $\sigma$) than any of the
giant ellipticals.
}
\label{fits}
\end{figure}

The means and standard deviations are displayed in Figure \ref{fits}.
Direct application of Eq.~(5) yields mean metallicities 
of $\langle Fe/H
\rangle_{blue} = -1.30 \pm 0.10$ and $\langle Fe/H \rangle_{red} =
-0.25 \pm 0.10$, along with mean intrinsic dispersions 
$\sigma$[Fe/H] $ = 0.32$ dex (blue) and 0.43 dex (red).  

For comparison, in Table 2 and Fig.~\ref{fits} we show the values
for the red and blue cluster populations in the Milky Way.  [Fe/H] data from
\citet{har96} were used to find the best-fit means and dispersions in
metallicity, and these were translated into equivalent $(B-I)$ values
through Eq.~(5).  For the Milky Way, the blue population is characterized
by $\langle$Fe/H$\rangle = -1.56\pm0.04$, $\sigma_{[Fe/H]} = 0.34\pm0.04$,
and the red population by
$\langle$Fe/H$\rangle = -0.55\pm0.04$, $\sigma_{[Fe/H]} = 0.16\pm0.03$.
The red populations in the giant ellipticals
are more metal-rich, and cover a very much broader range, than their counterpart in
the Milky Way. 
Furthermore, the mean abundance of the blue clusters in the BCGs is consistently 
$0.1 - 0.2$ dex more metal-rich than in the Milky Way.
As has also been discussed elsewhere \citep{strader04,wood05},
this comparison provides partial evidence that these BCGs cannot have
assembled {\sl only} from mergers of pre-existing galaxies the size
of the Milky Way or smaller, even with the addition of more metal-rich
clusters formed from gas in the mergers.
  
A correlation between the peak color of the blue and red modes and 
parent galaxy luminosity has been found \citep[see, e.g.][for
recent discussions]{forbes97,ff01,strader04} in
the sense that more luminous galaxies on the average have redder
GCs of both types.  However, all the galaxies in our sample are at
the high-luminosity end of this sequence, so the 
galaxy-to-galaxy differences in GCS color are expected to be small.
In fact, we find no evidence at all in our sample that the blue clusters
differ in mean color from one galaxy to the next:
the measured dispersion of 
$\pm 0.03$ mag in the mean color $(B-I)$ 
(Table 2) can be completely explained by the combination
of internal zeropoint scatter in the photometry and the external uncertainties
in the foreground reddenings (both of which are typically $\pm 0.02 $ mag).
Furthermore, their internal color range $\sigma_{B-I}$ has a galaxy-to-galaxy
dispersion of just $\pm0.02$ mag.

For the red clusters, the dispersion in mean color is $\pm 0.05$ mag, 
and the range in $\sigma_{B-I}$ is $\pm0.04$ mag, hinting at somewhat
larger intrinsic variation than the blue population.  Although our
sample of BCGs is small, the characteristics
of the red clusters do not obviously correlate with any other properties
of their parent galaxies, such as luminosity or surrounding environment.
As will be discussed later, if the red population formed primarily from
the last few major starbursts that built the majority of the galaxy, then
its characteristics could well be subject to large stochastic variations.

\section{Artificial-Star Tests}

To determine the effects of pure measurement scatter on the globular cluster
color distributions, we carried out a series of artificial-star tests on
the images.  These were done
in the usual way with {\sl daophot/addstar} by adding scaled PSFs at random locations
in the $B$ and $I$ images of each galaxy.  The added stars were defined to 
lie along vertical dispersionless sequences at $(B-I)_0 \simeq 1.65$ 
and $(B-I)_0 \simeq 2.05$, 
covering the range of magnitudes occupied by the real globular clusters in the image.
Typically we add 500 to 1500 artificial stars at a time; these do not add
noticeably to the crowding level of the frames, since all of them are
very uncrowded in any absolute sense.  We then measure the synthesized $B,I$ images
with exactly the same {\sl daophot} procedure as before, and find out how
many of the added objects are recovered and at what magnitudes.

\begin{figure}
\plotone{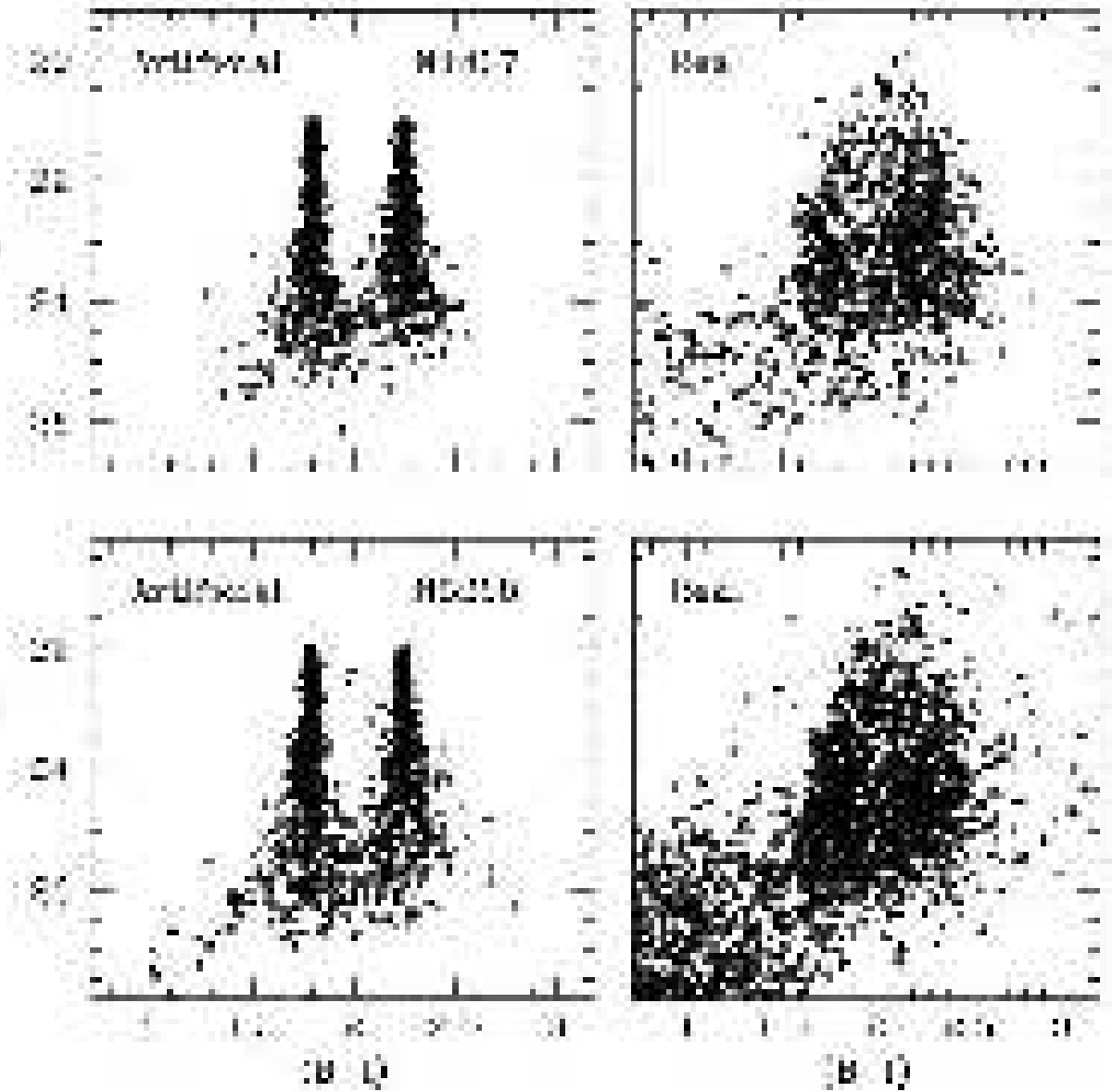}
\caption{Artificial {\sl addstar} tests for internal measurement errors of the
photometry.  {\sl Upper panels:}  Measured magnitudes and colors for the
artificial stars inserted into the NGC 1407 field, compared with the actual
measured data.  The added stars all fall along
lines exactly at $(B-I) = 1.80$ and 2.25, and are added in equal numbers per
unit magnitude.  No attempt has been made here to match the intrinsic
luminosity function of the real globular clusters.  {\sl Lower panels:}  The same
data tests for the NGC 3268 field.}
\label{fakecmd}
\end{figure}

Some sample results are shown in Figure \ref{fakecmd}, for NGC 1407 and 3268.
The results for the other fields look very similar, because the exposure times
were planned to yield very much the same photometric limits in 
{\sl absolute} magnitude $M_I$ in the globular cluster
luminosity function.  Since crowding effects are negligible, the
measurement scatter in the colors is generated almost entirely by the 
background image noise and photon statistics. No bias (systematic error)
is present at any magnitude level of interest, and for the brightest
$\sim 2 - 3$ mag of the globular cluster distribution, the measurement scatter
is far less than the {\sl observed} dispersions in color for both the blue
and red populations.  These tests provide strong evidence that we are fully
resolving the true color spreads of both the red and blue populations of clusters.

The artificial-star tests were also used to evaluate the completeness $f$
of the photometry.  We define $f$ as the number of artificial stars
inserted with {\sl addstar} in a given narrow magnitude range that were 
found and measured, divided by the number actually inserted in that
magnitude bin.  To be labelled as successfully ``found'', a star had
to be recovered in both $B$ and $I$.  Over the range of magnitudes that
we are using in this paper to discuss the metallicity distribution
(and excluding the innermost $5''$ regions around each galaxy center
as noted above),
our photometry is highly complete ($> 95$\%) and the MDFs unaffected
by bias.

\section{The Brightest Globular Clusters in the MDF:  A New Feature}

\begin{figure}
\plotone{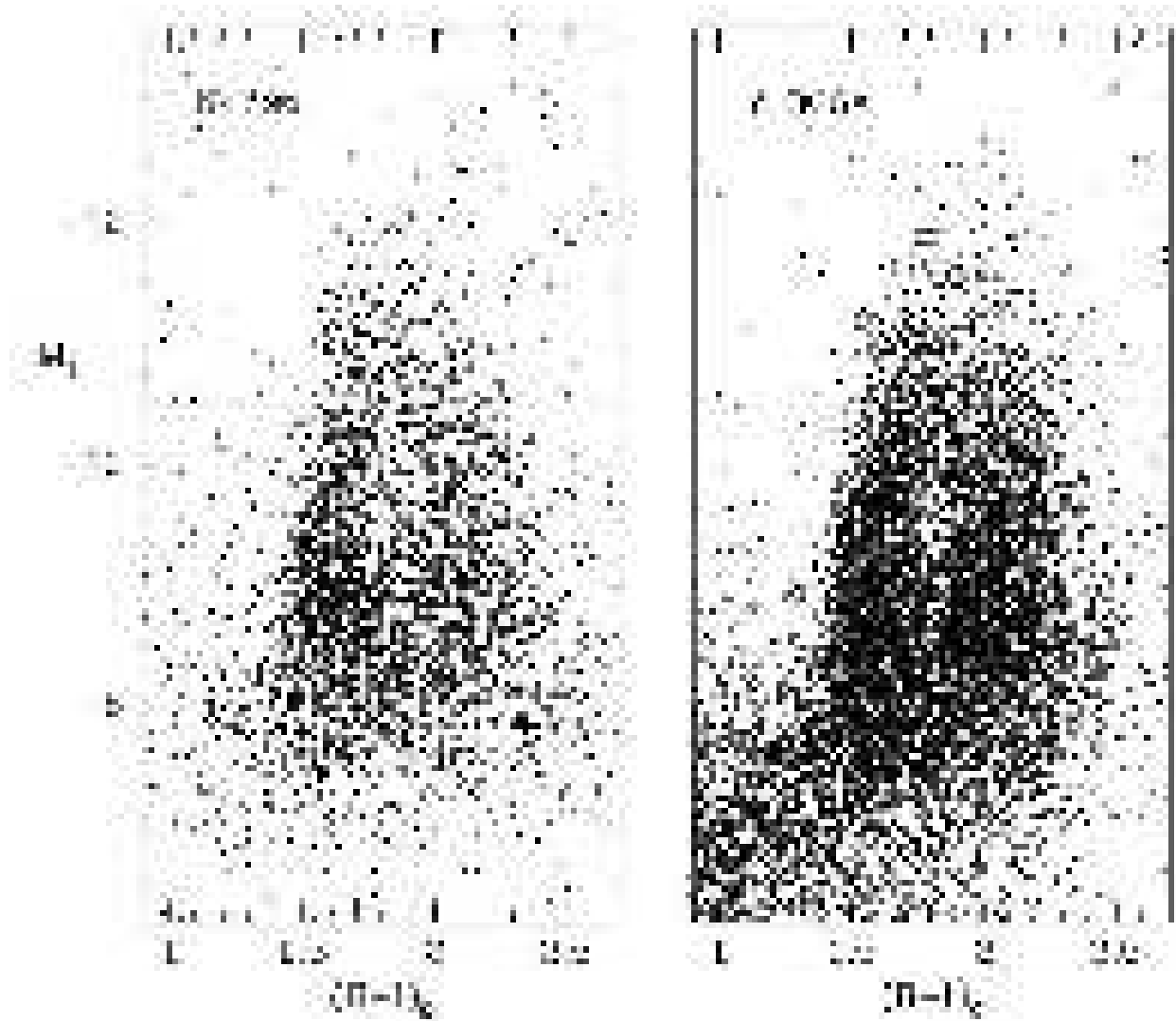}
\caption{{\sl Left panel:}  Expanded color-magnitude diagram for the GCS in
NGC 4696, the single largest cluster population in our survey.
The dashed line is at $M_I < -10.5$.  {\sl Right panel:}  Color-magnitude diagram
for the GCSs in the other 7 galaxies in our survey, normalized to absolute
magnitude and intrinsic color and combined.  
The three {\sl open stars} mark the locations of
massive Local Group globular clusters NGC 6715, $\omega$ Centauri, and (at top) M31-G1.}
\label{2cmd}
\end{figure}

While plotting up the color-magnitude array for NGC 4696, which is the galaxy
with the largest single cluster sample in our survey, we noticed
a new feature of the MDF:  for magnitudes
more luminous than $M_I \simeq -10.5$ (corresponding to $I < 23$), the 
distribution of cluster colors does not appear strongly bimodal, unlike
the very obvious bimodality at fainter levels.
The effect is somewhat as if the distinct blue and red modes that dominate at lower
luminosities simply merge together as they go up to higher luminosity.
In Figure \ref{2cmd} the
data are shown on a bigger scale, with the approximate dividing line
$M_I = -10.5$ below which bimodality is clearly evident.
This line is {\sl not} meant to imply an abrupt transition (indeed,
the trends shown in Fig.~\ref{all8_200} indicate a smooth, gradual
changeover), but only a guide to the point where this effect starts
to take place.

We then looked for the same feature in the other galaxies.  Though all the
rest have smaller observed sample sizes, the same characteristics can be seen,
and more prominently in the galaxies with larger measured GCS populations. The
composite color-magnitude distribution for the seven other galaxies
combined is also shown in Fig.~\ref{2cmd}, which 
comprises a sample twice as
large as in NGC 4696 alone.  The same feature shows up.
{\sl For the most luminous globular clusters, either 
the bimodality of the MDF disappears, or else the two modes
overlap so closely that they can no longer be cleanly distinguished.}

\begin{figure}
\plotone{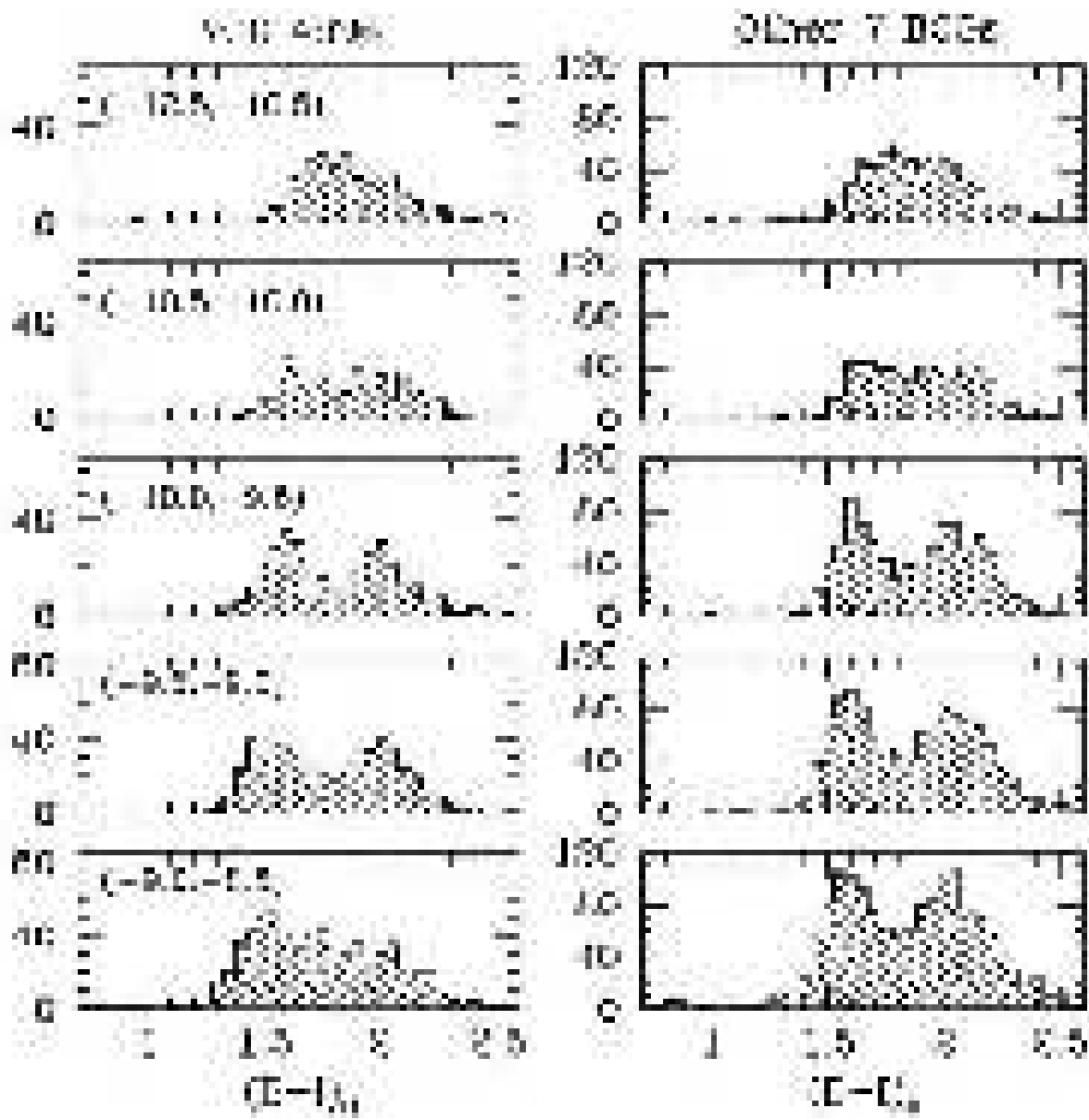}
\caption{{\sl Left panel:}  Histogram for the globular clusters in NGC 4696,
derived from the previous figure and 
divided into $M_I$ absolute magnitude bins as labelled.  {\sl Right panel:}  The same
distributions for the other seven BCGs combined.  
For $M_I < -10.5$ the clusters follow an increasingly unimodal color distribution.}
\label{maghisto}
\end{figure}

Color histograms for NGC 4696 and the other seven galaxies, subdivided
by magnitude, are in Figure \ref{maghisto}, showing the
trend for bimodality to weaken at higher luminosities.
The tests of the photometry described in the previous section
show that it is not the result of any purely internal
effects such as crowding, saturation, incompleteness, measurement bias, 
field contamination, or differences between PSF and aperture photometry.

\subsection{Statistical Tests}

Is this ``top end'' of the cluster distribution actually a broad, unimodal
distribution in color?  The tests of the KMM bimodal fitting on the composite
sample described previously
suggest that it is more likely to be the result of the two modes (blue
and red) that are very distinct at lower luminosities, but that converge at
higher ones.  But to investigate this further, we
have applied the same KMM mixture modelling tool as described earlier
to the {\sl individual} galaxies, in order to test how well the luminous
region ($M_I < -10.5$) can be matched by a single Gaussian distribution.
Our key numerical results are in Table \ref{brightKMM}.
In the Table, columns (2) and (3) give the probability $p$ that a single Gaussian
fits the MDF satisfactorily in the $M_I$ range $(-13.5, -10.5)$, followd
by the number of clusters in that luminosity bin. 
Columns (4) and (5) give the same numbers for a slightly fainter range
$(-9.8, -9.5)$ that has almost the same sample size as the top end, so that the
two probability values can be compared directly.  Finally, columns
(6) and (7) for completeness give these same numbers for the much broader
range $(-10.5, -9.0)$.  We conclude that for all the galaxies, the MDF is very clearly
bimodal in that fainter range.  Blank entries are ones for which
the KMM fit did not converge.

We look now particularly at the results for the bright end.  The smallest
samples (such as for NGC 7049 with 45 bright clusters, or NGC 3348 with 66)
return tolerably large $p-$values, indicating that a unimodal
Gaussian provides an adequate fit.  In general, however,
the probability $p$ becomes lower as sample size increases, which is
exactly the result expected if the underlying distribution requires
more than a single Gaussian to fit it.  
At the same time, for all the galaxies the $p-$value
decreases dramatically for the fainter bin (the second pair of 
numbers in the Table) compared with the
brighter bin, even with the same sample size.  
All these results are entirely consistent with the interpretation that the
MDF is intrinsically bimodal, but that the two modes do overlap
more closely at progressively higher luminosity.

As a concluding point of discussion, we return to the original
question:  is the MDF for the luminous ($M_I < -10.5$) clusters actually
unimodal?  It is trivially true, but nevertheless worth restating,
that the KMM fitting test does not answer that question.
Instead, it answers a much more restricted question:  it gives
the probability that a single Gaussian function matches the
distribution.  But there is no underlying physical reason why
the globular cluster MDF should obey a Gaussian or double-Gaussian
form:  different chemical evolution models and cluster
formation models can be found in the literature
that are all physically based, and that produce non-Gaussian but
unimodal MDFs.
An intrinsically unimodal distribution might be asymmetric,
or more sharply peaked than a Gaussian, or broader and flatter.
In all these other cases, any statistical fitting code that uses
Gaussian functions will require two or more such functions to
generate an adequate fit to the data 
even though the underlying shape of the actual MDF has a different nature.

As most previous authors have done, we regard the KMM 
Gaussian-based fits only as guides that provide useful numbers
to quantify the mean colors and intrinsic dispersions of the 
blue and red cluster populations.  
In our view, the main argument that the MDF has a bimodal form 
at all luminosity levels is
the gradual trend with $M_I$ that was presented in Section 3 above.
That is, the ``transition point'' at $M_I \simeq -10.5$ does not
mark any kind of abrupt change from bimodal to unimodal; it simply
marks the luminosity at which the two modes begin to overlap
strongly \citep[see also][]{strader05}.  We will quantify the global trend in color with
luminosity below, and suggest an interpretation for it.

\subsection{Is This a Universal Phenomenon?}

Viewed with the benefit of hindsight, exactly the same feature at the
bright end of the MDF -- a merging of the bimodal distribution into
a broad unimodal one -- can be seen in some other galaxies from
the data in the literature.
\citet{ostrov98} first remarked that the brightest clusters in NGC 1399
appeared to show a broad, unimodal color distribution in $(C-T_1)$,
splitting into a bimodal distribution fainter than $T_1 = 21$
(roughly equivalent to $M_R = -9.9$, very similar to the crossover level
we have identified).  \citet{dirsch03}, from a
more extensive photometric study of NGC 1399 again in $(C-T_1)$, 
concluded the same thing (see especially their Figure 11).  
In NGC 5128, a clear hint of this feature can be seen in the $(C-T_1)$
color-magnitude distribution shown by \citet{hhg04}; although the
cluster sample in this case is not large, it is biased strongly
towards the brightest objects by 
observational selection and so the top end is covered fairly well.
And in M87 a hint of this same feature may also be seen in
the WFPC2 observations of \citet{whit95} and \citet{larsen01}, though both the sample
size and metallicity sensitivity of the $(V-I)$ index used there
are too low to make it unambiguous.

Notably, all three of these other galaxies are BCGs:  M87 in Virgo, NGC 1399 in
Fornax, and NGC 5128 in the Centaurus group.  The question immediately
arises whether this intriguing characteristic of the MDF 
is peculiar to these central giant galaxies, and if so, why?  There is
not yet enough evidence to start answering this question.  Large samples
of clusters, and a metallicity-sensitive color index, are both needed
for this kind of MDF analysis to be done.  In other words, if the bright-end
``merging'' of the MDF modes is 
a universal feature of GCSs, the reason we see it prominently
in BCGs may simply be a result of the sheer size of the cluster populations
that they hold.  The bright-end clusters are rare to begin with, so that
in smaller galaxies with fewer clusters in total, it would not be possible
to subdivide the MDF by magnitude without losing out to
small-number statistics.    Many systems would have to be co-added to
regain statistical significance and look for the effect.

But some counterexamples to the simple population-size
explanation may already exist, such as in the Virgo
system NGC 4472 (M49), a giant E galaxy comparably luminous with M87
but not at the center of the entire Virgo cluster.  Two recent 
photometric studies with
wide-field coverage and large statistical samples 
of the GCS \citep{gei96,rhode01}
plainly reveal the metal-rich and metal-poor modes in the color-magnitude
distributions of the clusters, and yet both studies indicate that
the two modes remain fairly distinct at the high-luminosity end.
Thus a {\sl very} tentative suggestion from this one comparison is
that the phenomenon we have isolated may indeed be associated most strongly with cluster
formation in the deep potential wells of the brightest central giant galaxies.
This argument provides additional reasons 
to look for the phenomenon in small galaxies of other types.
An excellent prospect for doing such a test would be with the
ACS Virgo Cluster Survey \citep{cote04} of a hundred normal galaxies
covering a wide range of luminosities.

Recently \citet{strader05} have discussed ACS $g'z'$ photometry of the 
cluster populations around several galaxies including the Virgo giants
M87, NGC 4472, and NGC 4649.  The $(g'-z')$ color index has a metallicity
sensitivity very similar to $(B-I)$, allowing the blue and red
MDF modes to be clearly identified in their GCS populations.  They find that for M87, and 
possibly also in NGC 4649, the blue-cluster mode becomes progressively
redder at higher luminosity, basically the same trend we find here.
In NGC 4472, they find no such trend, in agreement with the previous
studies cited above.  The advantage of their dataset is that it is
internally homogeneous and contains a wide variety of E galaxies;
their results are consistent with our suggestion that 
the color/luminosity phenomenon is associated with central giant 
ellipticals.  

We note that \citet{ostrov98} and  \citet{dirsch03} described
the bright clusters in NGC 1399 as displaying a unimodal color distribution, whereas
\citet{strader05} state that the blue clusters in M87 follow a color/luminosity
trend.  Our view is that these two descriptions are {\sl both} results of
one and the same phenomenon.

\subsection{Mass Fractions}

In our combined material for all eight galaxies, there are almost 4600
globular clusters brighter than $M_I = -9.0$ (Table \ref{field}).
If as noted earlier the GCLF has a normal Gaussian-like shape
with turnover at $M_I = -8.4$ and dispersion $\sigma \simeq 1.4$ magnitudes
\citep{h01}, then the total
GCS population we are drawing from in all eight galaxies, and within
the ACS field of view, is nearly 15000 clusters.  A total of 970 of
them, or just 6.5\%, are brighter than $M_I = -10.5$.
But even though the clusters brighter than this dividing line
are rare, they contribute a large fraction of the total
{\sl mass} in the system:  
$M_I = -10.5$ corresponds to $L = 5 \times 10^5 L_{\odot}$,
or $M = 1.5 \times 10^6 M_{\odot}$ for a visual mass-to-light ratio of 3.
With the GCLF parameters as noted above, 
and with the assumption that the cluster
mass-to-light ratio is roughly independent of luminosity, then the
clusters brighter than $M_I = -10.5$ make up a full 40\% of 
the total mass that is now in the GCS.  In other words, they are major
players in the formation history of star clusters,
using up a significant fraction of the gas that was eventually incorporated
into bound clusters during the galaxy formation era.\footnote{The 40\% mass
ratio is, obviously, the fraction of globular cluster mass {\sl now} taken up
by these most luminous clusters, after more than 10 Gy of dynamic evolution
within the potential wells of their parent galaxies.  We would ideally
like to know instead the fraction of
cluster mass that originally went into these biggest objects.
This initial ratio should be smaller, because the lower-mass objects are
preferentially destroyed or eroded by evaporation and tidal shocking.
Recent simulations accounting for this 
\citep[e.g.][]{fall01,ves03} indicate that these most massive
clusters would, however, still take up $\sim 20$ percent or more of
the GCS mass at earlier times.}

In the Milky Way \citep{har96}, there are two globular clusters known which
clearly belong to this high-mass regime:  $\omega$ Centauri at $1.0 \times 10^6 L_{\odot}$
and NGC 6715 at $0.8 \times 10^6 L_{\odot}$.  A few more are known in M31, the
most notable of which is M31-G1 \citep{meylan01} at $2.3 \times 10^6 L_{\odot}$.
The positions of these local examples are shown for comparison 
in Fig.~\ref{2cmd}.  In our entire sample, the brightest
single objects can only be identified statistically since the numbers
trail off indistinctly with luminosity (see again Fig.~\ref{2cmd}), but
the top end appears to be near $M_I \simeq -12.6$, equivalent to more than 12 million
Solar masses.  M31-G1 particularly is near the very maximum 
of the mass range that globular clusters reach in the real universe.

\subsection{Image Morphologies and Spatial Distributions}

An immediate question is whether or not the brightest GCs we
see around our target galaxies are somehow
different in any obvious way from the fainter ``normal'' ones, possibly
indicating a different type of origin.  
For example, the Ultra-Compact Dwarfs (UCDs) \citep{phil01,drink03} 
have comparable luminosities to the brightest GCs, small linear
sizes, and could be the remnant nuclei of disrupted dE's.  Such
objects could help populate the bright
end of our GCS luminosity distribution \citep[see][for a recent
example]{mie04}.  
Another recently discovered type of cluster-like object is W3 in NGC 7252 
\citep{maraston04}, which has both a remarkably high mass ($8 \times 10^7 M_{\odot}$)
and an effective radius (18 pc) several times larger than a typical globular cluster.

\begin{figure}
\plotone{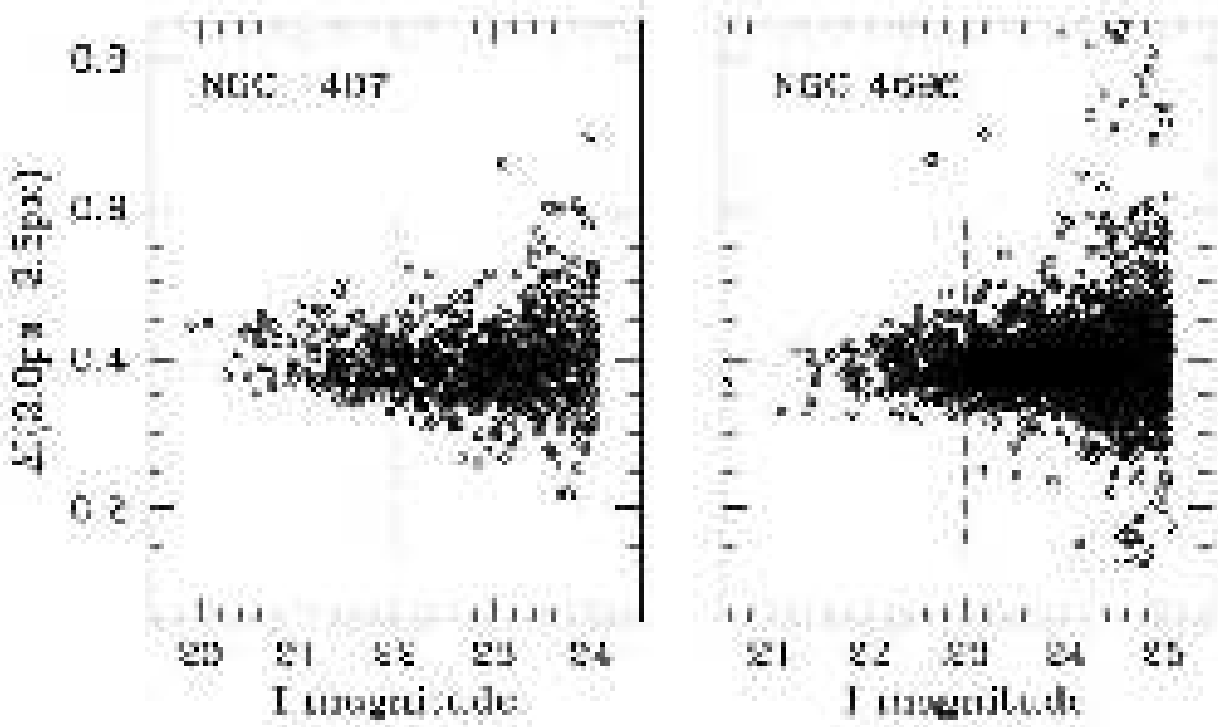}
\caption{Tests for nonstellarity of the bright
globular clusters.  The {\sl left panel} shows the magnitude difference
$\Delta I$ between a 2-px and 3.5-px aperture radius versus $I$ magnitude
for the globular clusters around NGC 1407, the nearest galaxy in our
survey.  The horizontal line shows the mean value for starlike objects
(ones matching the PSF profile accurately), while the vertical dashed
line shows the approximate dividing line at $M_I = -10.5$ from Figure 10.
Ones fainter than $I \simeq 22$ define a bimodal MDF, while brighter
ones define a unimodal MDF.  Some of the clusters in the brighter
part are visibly nonstellar, scattering up to higher $\Delta I$.
The {\sl right panel} shows the same
distribution for NGC 4696, a galaxy twice as far away (see text).}
\label{fuzzyap}
\end{figure}

If any of the brightest GCs are of this type,
then they might have more extended structures than normal clusters.
We have looked for evidence of this effect.
A simple tracer of
nonstellarity is the magnitude difference between two small
apertures of different radii; two examples are shown in Figure \ref{fuzzyap}.  
Here the magnitude difference in $I$ between a 2-pixel-radius
aperture and a 3.5-pixel aperture is plotted against
magnitude.  Nonstellar objects will have more extended wings than
the point spread function and thus sit above the main distribution
defined by objects that match the PSF well.  

\begin{figure}
\plotone{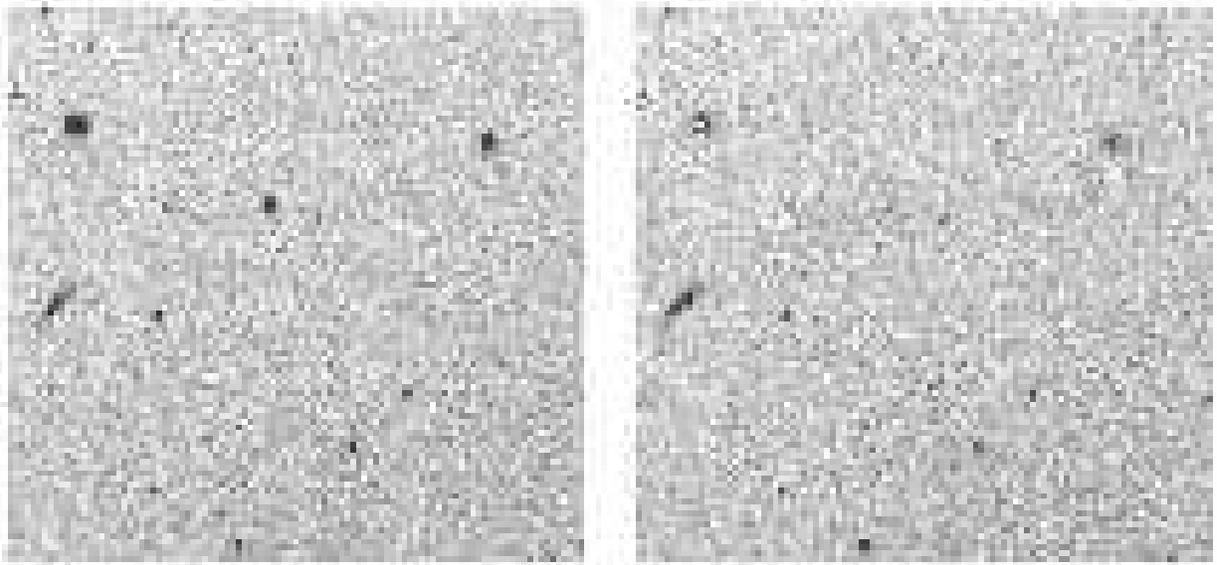}
\caption{{\sl Left panel:}  A $22''$ wide segment of the halo region around
NGC 1407, the nearest BCG in our survey.  The three bright objects
in the top half of the image are globular clusters in the
``unimodal'', bright-end part of the metallicity distribution function.
{\sl Right panel:} The same field after the point spread function
has been subtracted from the profiles of the three objects.  Note that
the middle one is nearly starlike and subtracts 
cleanly, but the other two have left residual
faint envelopes.  The fainter objects in the lower half of the frame
are all starlike.}
\label{fuzzypic}
\end{figure}

For NGC 4696, at a distance of 42 Mpc, Fig.~\ref{fuzzyap} shows
only the horizontal starlike sequence, with no strong evidence for any 
very extended objects.  At this distance, the FWHM of the
PSF (2.2 px) corresponds to a diameter of 25 parsecs 
(for comparison, $\omega$ Cen has a half-light diameter of 13 pc and
the UCDs have sizes similar to this or a bit larger).
Most of the galaxies in our list are at similar distances (Table 1) and
give similar results.
However, for the closest galaxy in our survey, NGC 1407 at 
d = 23 Mpc, clusters the size of $\omega$ Cen or bigger would
subtend half-light diameters of 2.2 px and thus should be
marginally resolved.  In fact, we find a
population of about 20 to 30 bright objects around
NGC 1407 that are slightly
more extended than the starlike sequence:  in Fig.~\ref{fuzzyap},
these are the ones with $\Delta I \geq 0.4$, $I < 21.5$.
Inspection of the images shows that faint extended envelopes are present
around several of them; for
the sake of visual comparison, we show three of these in
Figure \ref{fuzzypic}.  Such objects have morphologies
very similar to the way $\omega$ Cen would look at that distance.

At the same time, few if any objects that are significantly {\sl more}
extended in morphology show up in our data.  Although this should in large part
be a selection effect (faint objects with extremely different sizes than the image
PSF will tend to be rejected in the photometry), it verifies that essentially
all the objects in our samples can be considered as {\sl bona fide} 
globular clusters of the range of properties we have already seen within the Local Group.

\begin{figure}
\plotone{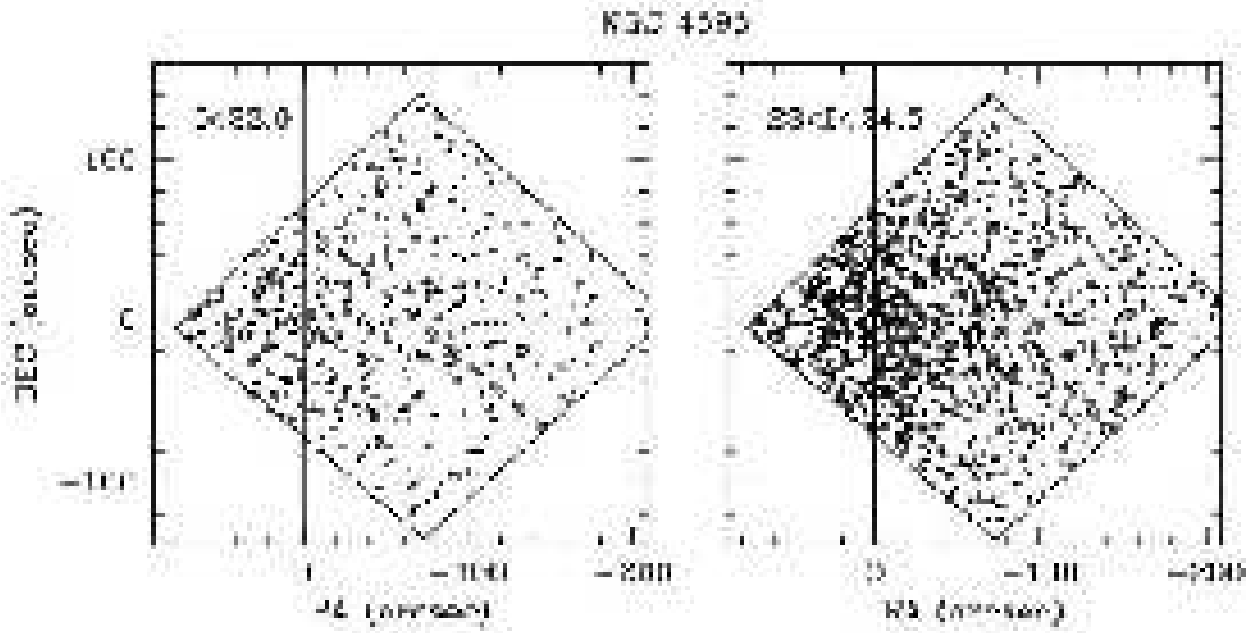}
\caption{{\sl Left panel:}  Locations of the brightest globular
clusters around NGC 4696.  These objects follow the broad unimodal MDF 
shown in Fig.~7 above.  The galaxy center is marked by the crosshairs and a $5''$
circle around the nucleus. The quadrangle, about $3\farcm4$ on a side,
marks the borders of the Multidrizzled
ACS/WFC frame, and the image is oriented with North up and East to the left.  
The dashed lines show the dead space
between the two WFC CCD detectors.  {\sl Right panel:}  Spatial distribution for
fainter globular clusters in NGC 4696, which follow a normal bimodal MDF.
Both bright and faint clusters are concentrated to the galaxy center.}
\label{xyplot}
\end{figure}

\begin{figure}
\plotone{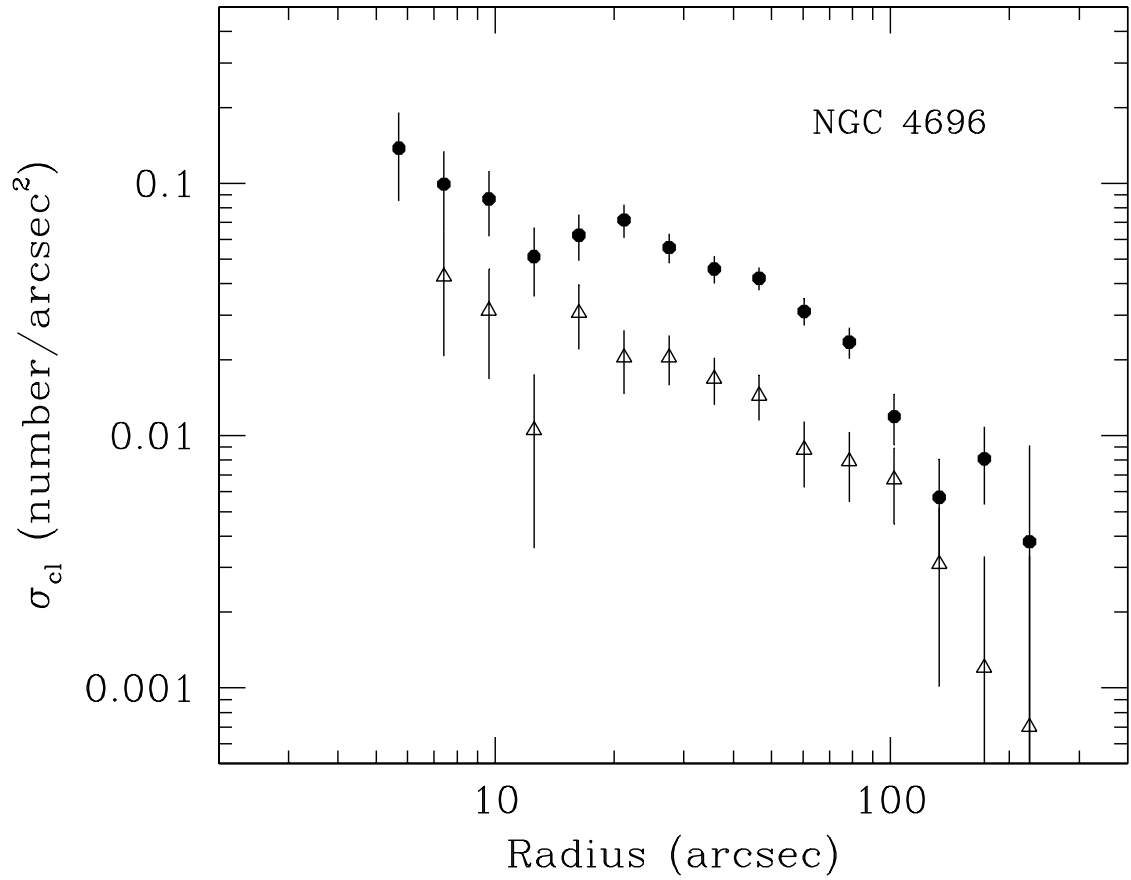}
\caption{Spatial distribution of the globular cluster system in NGC 4696.
The number of measured objects per unit area in the NGC 4696 ACS field
is plotted against projected galactocentric
distance.  Open triangles show the bright clusters $(I < 23)$,
while solid symbols are fainter ones $(23 < I < 24.5)$.  For these relatively
bright magnitude intervals, completeness corrections and field contamination
are negligible (see text).
}
\label{radplot}
\end{figure}

In addition, as far as we can tell, the luminous ``unimodal'' clusters do not have
any different spatial distribution than the less luminous clusters.  
An example is shown in Figure \ref{xyplot}, 
where the locations of
the brightest clusters in NGC 4696 are plotted in comparison with
a large sample of the fainter ones.  No differences are evident; the
brightest ones are found everywhere through the halo but with the same
degree of central concentration as the others.  More detailed tests of
the radial profiles (shown in Figure \ref{radplot}) indicate no
quantitative difference in the slope 
$\Delta {\rm log} \sigma_{cl} / \Delta {\rm log} r$.  
The other galaxies, though with smaller
statistical samples to work with than NGC 4696, exhibit no different pattern.
This evidence indicates that the broad spread of colors at the
high-luminosity end is not associated in some way with a particular type
of cluster formation (say) in the inner bulge region or the outer halo.

In summary, aside from a few objects with slightly more extended structures,
the brightest clusters show no additional distinguishing features
from the rest of the globular cluster population.  
We conclude that their color distribution must then correlate in some
more direct way with cluster luminosity and (thus) mass.

\subsection{Connections to dE Nuclei}

Another special group of objects strongly resembling the globular
clusters in luminosity, size, and (probably) age are the nuclei of dwarf
elliptical galaxies.  The massive cluster
NGC 6715 forms the nucleus of the Sagittarius dwarf elliptical
\citep{lay00}, and
the extended structure and multicomponent population within 
$\omega$ Centauri have raised suggestions that it might be the
relic nucleus of a former dE satellite of the Milky Way
\citep[e.g.][]{maj00,mcw05}.  Similar ideas have been raised for M31-G1 
\citep{meylan01}.  Could the luminous cluster population
be a collection of dE,N nuclei accreted long ago and stripped
of their envelopes?  This discussion can be traced back to early ideas
by \citet{zin88} and \citet{free93} and has been explored in a variety of
ways since.  The sheer numbers of such
objects do not appear to present an obstacle by themselves, 
since in normal hierarchical-merging galaxy formation models it
takes thousands of dwarf-sized pregalactic clouds\footnote{In this paper, 
we use the term ``pregalactic clouds'' to refer to the population
of gaseous, dwarf-galaxy-sized, protogalactic
disks that started merging
to form bigger galaxies; that is, it refers to the epoch just
at the beginning of galaxy formation and not still earlier
cosmological epochs.} to eventually
accumulate into one giant elliptical. If only a fraction of these
happened to contain very massive globular-cluster-like nuclei, they could account
for the few hundred that we see in our combined sample.
Furthermore, extensive quantitative models have been developed to explore
the possibility that many or most of the low-metallicity globular
clusters in major galaxies might have been accreted from
dwarf satellite galaxies \citep{cot00}.  In this view, a bimodal
MDF is not a required outcome, but rather 
a statistical result of the particular mass spectrum of accreted satellites,
so a broad unimodal MDF is a definite possibility.  (It is not at all clear,
however, how this model can give rise to an MDF within one galaxy
that is simultaneously bimodal at lower
luminosities and yet unimodal at higher luminosities.)

Objects such as UCDs and
the nuclei of nucleated-dwarf dE,N galaxies are quite hard to tell
apart from traditional globular clusters if we
have data in hand that consist only of luminosities, colors, and upper
limits on their characteristic sizes.  However, if we cannot individually
identify clusters that {\sl were} once dE,N nuclei, we can at least compare
our BCG globular cluster population with the nuclei that are {\sl now}
in dE,N dwarfs.  Fortunately, a comprehensive and homogeneous photometric
study of dE,N nuclei from HST/WFPC2 imaging exists \citep{lotz04} that 
can be used for this purpose.

\begin{figure}
\plotone{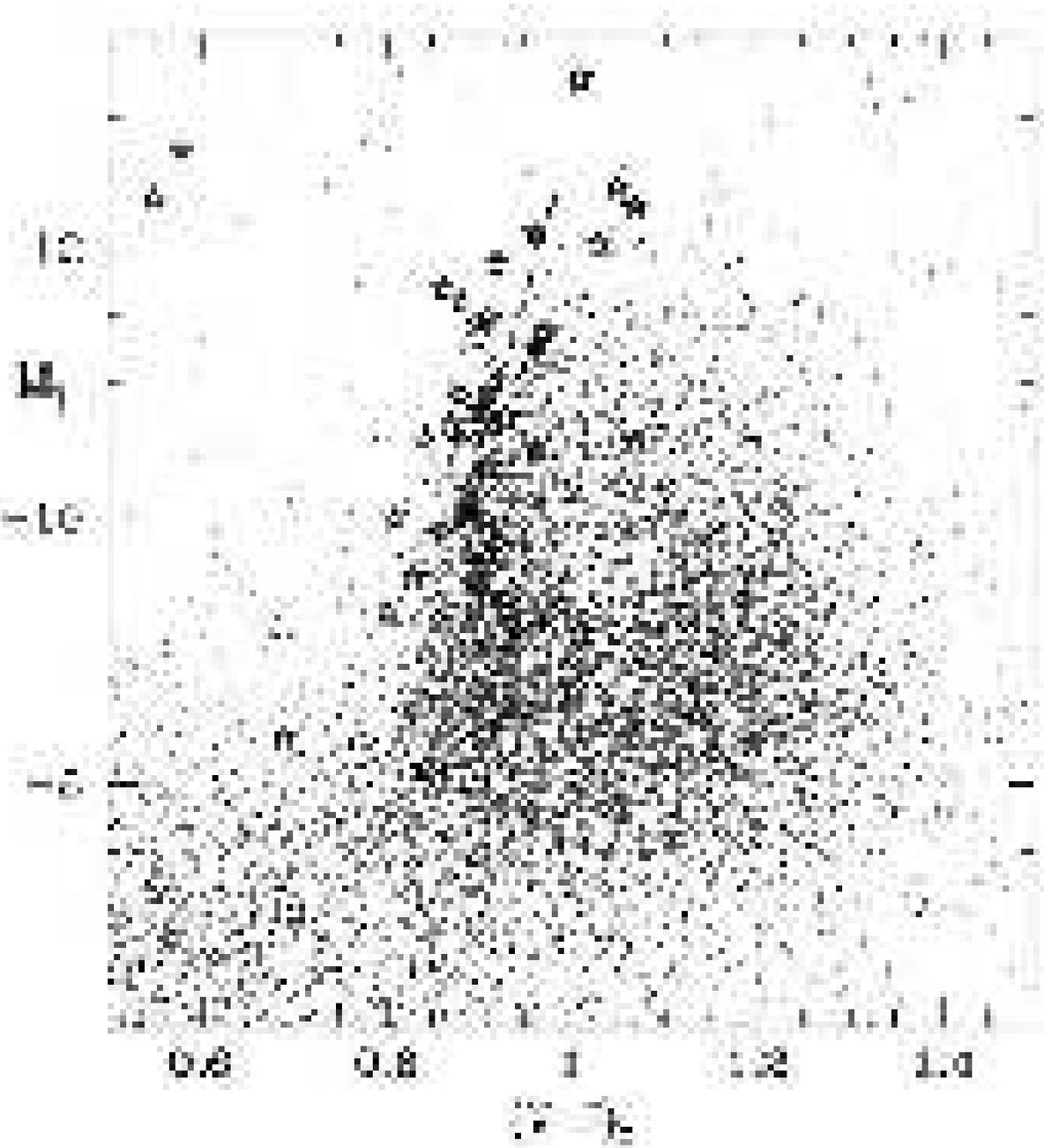}
\caption{Composite color-magnitude diagram for the clusters in all
eight of our BCGs, where our measured $(B-I)$ colors have been transformed
to equivalent $(V-I)_0$.  The large open stars show the locations
of the nuclei of {\sl nucleated dwarf} dE,N galaxies, from Lotz
et al. (2004).  The dashed line is explained in the text.
}
\label{cmd_de}
\end{figure}

\begin{figure}
\plotone{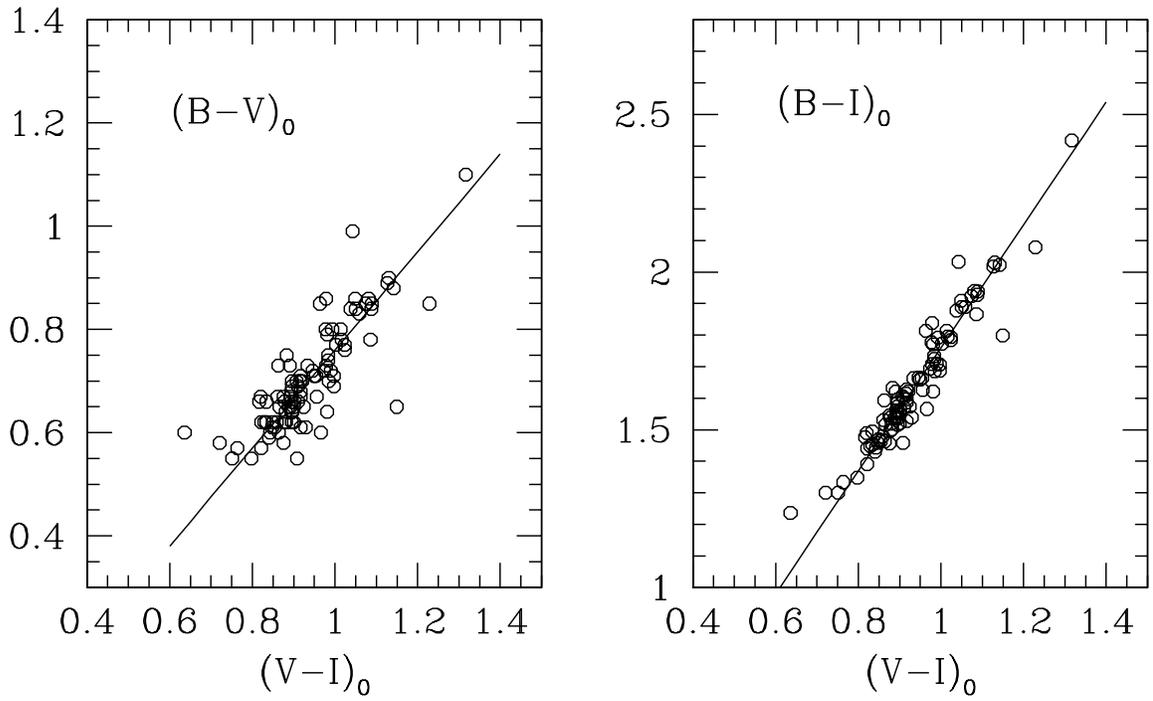}
\caption{Intrinsic colors for globular clusters in the Milky Way,
with catalog data from Harris (1996). The left panel shows $(B-V)_0$
versus $(V-I)_0$, and the right panel $(B-I)_0$ versus $(V-I)_0$.
Equations for the fitted lines are given in the text.
}
\label{2color}
\end{figure}

We show this comparison in Figure \ref{cmd_de}. 
Here, the data from all eight of our BCG fields are combined together, and
overplotted with the photometry for the dE,N nuclei in the Virgo and Fornax
groups, with data taken from the catalog of \citet{lotz04}.
To the Lotz et al. measurements we have added three more dE,N objects
for which luminosities and colors are available:
NGC 6715 in the Sagittarius dwarf (data from Harris 1996), and the nuclei 
of NGC 5206 \citep{cal87} and NGC 3115B \citep{dur96b}.
To normalize the colors of all these objects to the $(V-I)$ scale used
by Lotz et al., we have converted our $(B-I)$ measurements and the
other $(B-V)$ data in the literature through empirical relations shown
in Figure \ref{2color}, defined by the Milky Way globular clusters.
The lines shown in the Figure are given by
\begin{eqnarray}
(B-V)_0 \, = \, 0.96 (V-I)_0 - 0.19, \\
(B-I)_0 \, = \, 1.96 (V-I)_0 - 0.19.
\end{eqnarray}

Aside from a couple of very blue nuclei (interpreted as younger,
more recently formed clumps; see Lotz et al. 2004), the nuclei define
a moderately broad sequence over a luminosity range $M_I < -9$ similar to
the traditional bright globular clusters.  The very most luminous nuclei extend
up to $M_I \sim -12.5$, as does the top end of the globular cluster
population.  However, the most striking feature of the dE,N sequence is that it lies  
entirely along the blue side of the globular cluster distribution;
essentially none of them fall within the red-cluster population.  
We can think of no observational selection effect that would prevent slightly
redder nuclei from being detected, and since the photometry is based on
HST/WFPC2 images, the nuclei can be well separated from contamination
by the surrounding dE envelopes.
Lotz et al. note that the colors of the nuclei are consistently bluer
than those of the dE galaxies they are embedded in, and thus by assumption
should be either old and metal-poor, or (if they are actually the
same metallicity as their host dE's) much younger.

At a finer level
of detail, the dE,N sequence is bluer by $\sim 0.05 - 0.1$ mag 
than the center of the blue-cluster sequence.  It is not clear if this offset in color 
is significant:  it may at least partly arise from
transformation errors between one color index and another, or the
combined zeropoint errors in the WFPC2 and ACS photometric data.
Their location would, however, be consistent with the interpretation
that most of the nuclei are old and metal-poor and thus acted as the first
``seed'' structures formed in their host dwarfs \cite[see][]{lotz04,dur96a}.

Regardless of how the dE,N nuclei are interpreted, it seems clear that
many of the luminous globular clusters in our BCG sample have a different origin.
Although dE,N nuclei could provide the source for many of the luminous
blue clusters, most of 
the globular clusters at all luminosities are quite a bit redder than
the nuclei.  The simplest interpretation is most of them
are not the relics of dE,N dwarfs that were
accreted by the BCGs at some later time long after the major formation
stage.  An alternate, albeit more complex, interpretation would be
that the 
dE,N dwarfs still remaining today (such as the Virgo and Fornax
members used above) are preferentially more metal-poor and isolated
than the ones accreted by the BCGs at earlier times.  
But we believe it is highly unlikely that dynamical processes could have
selectively removed all of the hypothetical dE,N galaxies with 
intermediate- or high-metallicity nuclei while leaving behind only a few
with metal-poor ones.

Another interpretation that has been offered for the UCDs 
and objects like NGC7252-W3
is that they are merged products of several young massive star clusters
that formed from cluster complexes in major starbursts
\citep{fell02}.  This mechanism also appears unlikely to explain the
luminous population we see:  as discussed by
\citet{fell02}, such merger products would have very 
large effective radii unlike globular clusters, and most
of them would be expected to be metal-rich if they formed during the
late major starbursts.

Our general conclusion from these comparisons is that some of the
luminous blue clusters could well be accreted dE,N nuclei, but
that the red cluster population is more likely to have a different
origin as normal globular clusters.

\section{Bimodality Model Fits:  A Mass-Metallicity Relation}

To test in more detail what happens to the two obvious ``modes'' (blue and red) as a
function of cluster luminosity, we carried out a series of 
tests where the MDF is modelled, as before, by a combination of two Gaussian distributions.
We allowed both the dispersion and mean color to be solved for.  In each galaxy
we grouped the MDF into magnitude bins containing 200 objects per bin, so
that each bin would have similar statistical uncertainties.

\begin{figure}
\epsscale{0.9}
\plotone{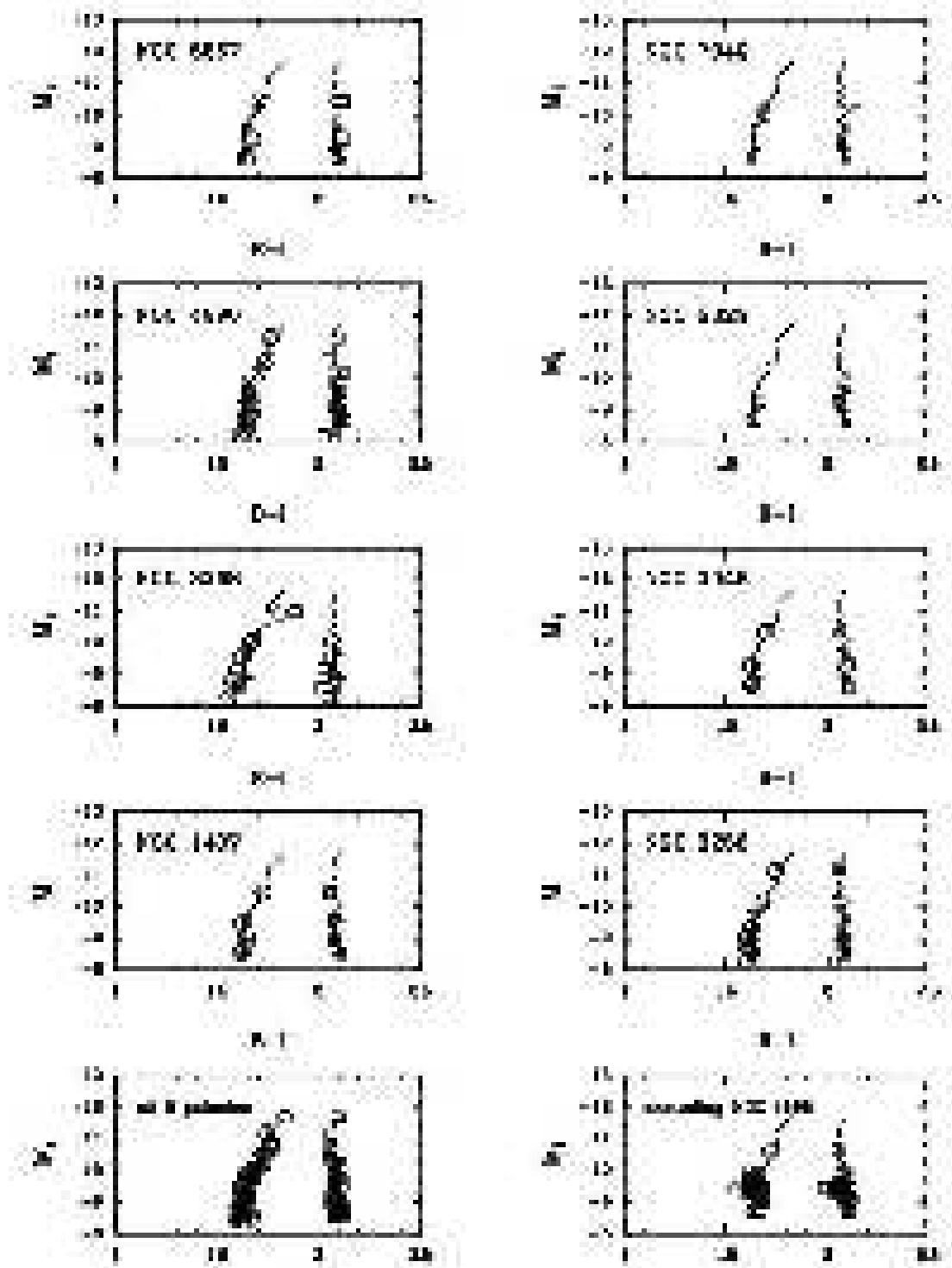}
\caption{Mean colors for the individual GCSs in the eight BCGs.  In each
panel, the clusters are subdivided into magnitude bins of 200 objects per bin.  
The KMM mixture modelling code is used to fit two Gaussians to the color
distribution in each bin, with resulting mean colors shown by the
open circles (metal-poor clusters) and open squares (metal-rich clusters)
in each panel.  The graph at {\sl lower left} shows the
composite population for all eight galaxies; the jagged line through
its set of points is reproduced in all the other panels for comparison.
The graph at {\sl lower right} shows the mean lines for seven galaxies
excluding NGC 4696.
}
\label{10panel}
\end{figure}

In Section 3 above, we described the fits to the composite 8-galaxy sample.
Here, we show the results for the individual galaxies, displayed in
Figure \ref{10panel}.  The lower left
panel is our previous composite fit for all eight galaxies combined and shows the
trend of mean color with absolute magnitude $M_I$ (traced by the two jagged
lines that pass through the mean points).  These lines are reproduced in
each of the other panels for comparison.  

Of the principal results emerging from these fits, we note first that since 
each mean point is independent of the others, the very small offsets
between the individual bin points and the mean lines
in Fig.~\ref{10panel} demonstrate that all the MDFs 
exhibit bimodality with a high degree of reliability, and with 
nearly identical patterns in the different galaxies.

Secondly, the mean color of the {\sl red} mode 
shows no significant change with luminosity.  That is, there appears to be no
mass-metallicity trend for the more enriched clusters that are the most
likely to have formed in the major series of starbursts that built the
main body of the parent galaxy.

Thirdly, the KMM solutions for the blue-cluster sequence suggest
a correlation between color and luminosity beginning at about   
$M_I \simeq  -9.5$.  The effect is most noticeable for
the galaxies with the {\sl largest} observed cluster populations
(namely NGC 1407, 3258, 3268, 4696) that carry the most weight
in the definition of the mean lines in Fig.~\ref{10panel}.
The other four (NGC 3348, 5322, 5557, 7049)
do not disagree with the same trend, but they do not carry enough
statistical weight by themselves to say whether the
same trend exists within them individually.
Although this correlation may at least partly be 
bound up with our lingering concerns about
photometric calibration and the small differences between psf-based
and aperture-based photometry, we find that it is present in
all our galaxies (see again Fig.~10b) regardless
of which type of data we use, either aperture or PSF-fitting.

\begin{figure}
\epsscale{1.0}
\plotone{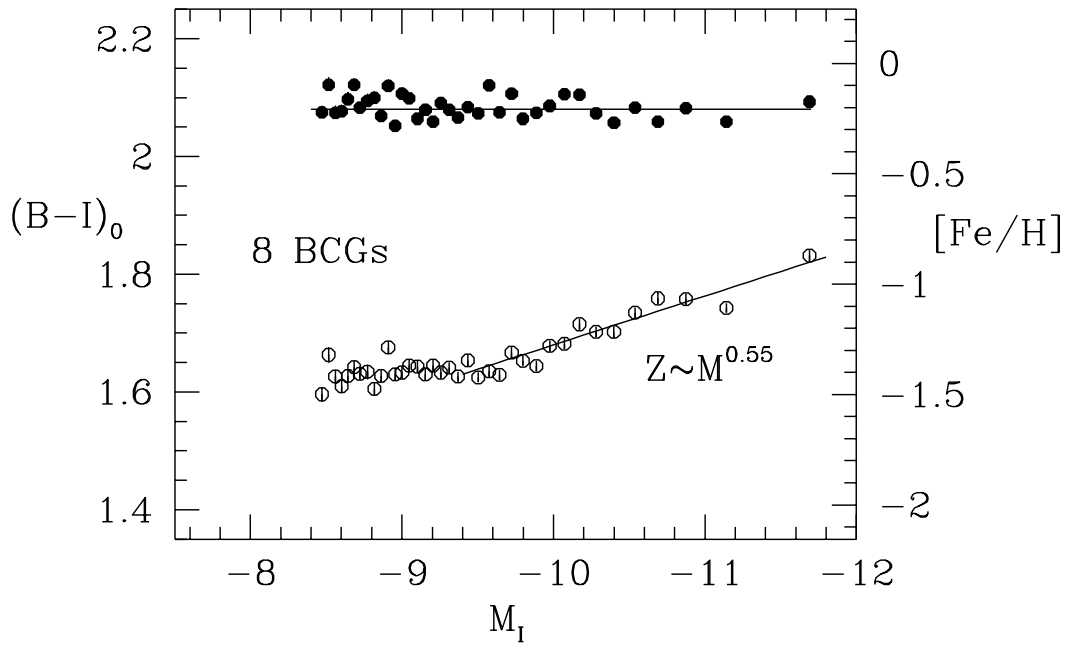}
\caption{Mean intrinsic colors of the globular cluster populations as a
function of luminosity, from the combined sample of eight BCGs.  Metal-rich
clusters are the solid symbols and metal-poor ones the open symbols.  
Each dot represents the mean of 200 objects; the uncertainty of each mean
is similar to the size of the symbol.  For the metal-rich clusters, the
mean color is  constant with luminosity at 
$\langle B-I \rangle_0 = 2.08$, equivalent to log $(Z/Z_{\odot}) \simeq -0.21$.
For the metal-poor clusters, objects brighter than $M_I = -9.5$ follow
a scaling with mass given by log $(Z/Z_{\odot}) = -5.0 + 0.55$ log $(M/M_{\odot})$
(shown by the sloped line).  
}
\label{zscale}
\end{figure}

Since $(B-I)$ depends linearly on [Fe/H], 
and absolute magnitude varies logarithmically with luminosity (hence mass),
the color/magnitude trend turns directly into a simple power-law scaling
of heavy-element abundance $Z$ with cluster mass $M$.  The result for the
combined sample is displayed in Figure \ref{zscale}, where we have
adopted as before a mean mass-to-light ratio $(M/L)_V \simeq 3$.  The
ratio $(M/L)$ is difficult to measure accurately and 
is likely to differ somewhat between 
clusters \citep[e.g.][]{meylan95,mcl00,baum03,pas04,mar04}
but we assume it to be roughly invariant with luminosity.  
For the blue cluster population, we then find 
\begin{equation}
{Z \over Z_{\odot}}(GC) \, \simeq \, 10^{-5} ({M \over M_{\odot}})^{0.55}
\end{equation}
for masses higher than about $6\times 10^5 M_{\odot}$ 
(equivalent to $M_I \simeq -9.5$, which is about
a magnitude brighter than the GCLF turnover).  

The dE,N nuclei discussed above follow a very similar trend of mean color with
luminosity, again starting at a similar luminosity level $M_I \simeq 10$.
This correlation is shown as the dashed line in Fig.~\ref{cmd_de}, which 
corresponds to a metallicity scaling $Z \sim M^{0.55}$.  This comparison
reinforces the idea
that the nuclei are structurally similar entities to the bright, low-metallicity
globular clusters.

Although too much weight
should not be placed on the datapoints at lower luminosities, where the
photometry becomes more uncertain and more affected by field contamination,
it does not seem likely that the same power-law scaling continues downward to
lower masses.\footnote{In this respect we disagree with the trend proposed for
M87 by \citet{strader05}.  They impose a simple linear solution of mean $(g'-z')$
versus magnitude extending over all luminosities.  Their actual color-magnitude graph,
however, suggests that the lower-luminosity clusters have a constant mean color
that does not follow their linear solution.}
The main reason for this suggestion is that on external grounds
we expect the ``bottom end'' in metallicity for the metal-poor clusters to
be at a mean [Fe/H] $\simeq -1.6$, where we find the blue mode
for clusters in the Milky Way and in a wide range of dwarf ellipticals 
\citep{hh01,burg01,lotz04,wood05}, and near where the low-luminosity
end of our BCG sequence lies.  

\section{Metallicity Dispersions}

\begin{figure}
\plotone{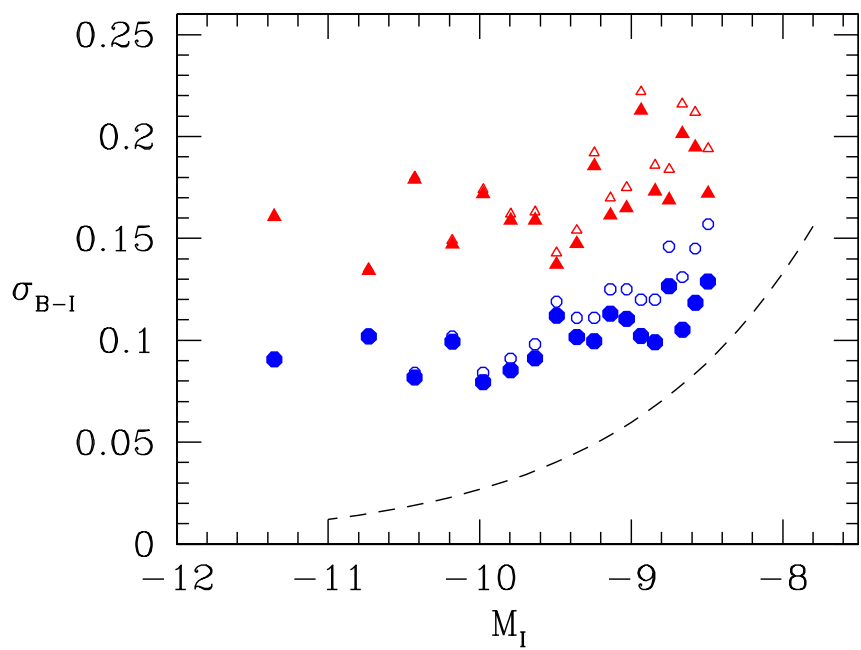}
\caption{Measured dispersions of the blue and red cluster MDFs.
Here $\sigma_{B-I}$, the rms dispersion of each mode as determined from
the double-Gaussian fits described in the text, is plotted versus $I$ magnitude:
small open circles are the metal-poor clusters, small open triangles the
metal-rich clusters.  The dashed line shows the average photometric measurement
uncertainty $\sigma_{phot}$; when this is subtracted in quadrature from
$\sigma_{B-I}$, the resulting estimate of the intrinsic metallicity
dispersion of each mode is shown with the larger solid symbols.
}
\label{dispersion}
\end{figure}

The intrinsic dispersions of the red and blue modes are 
well resolved in our data, and emerge more clearly here than in
most previous studies.  They can therefore be used to estimate the internal range
of abundances of each type of cluster.  
The internal dispersions are
noticeably different in the two modes, 
averaging $\langle \sigma_{B-I} \rangle = 0.099 \pm 0.007$
for the blue mode and $0.166 \pm 0.014$ for the red mode (Table \ref{bimodal}).
In Figure \ref{dispersion}, we show
$\sigma(B-I)$ as deduced directly from the KMM two-Gaussian model fits, plotted
as a function of luminosity $M_I$.  The directly observed dispersions 
grow progressively larger
for $M_I > -9.5$, but this is at least partly a result of the increasing
field contamination as well as the
photometric measurement uncertainty (shown as the dashed line in the figure; since
the photometric limits in $M_I$ are very similar from one galaxy to the next,
we show only the mean line).  A better estimate of the residual intrinsic dispersion 
is $\sigma_0 = (\sigma_{B-I}^2 - \sigma_{phot}^2)^{1/2}$, where the main effect
of at least the photometric scatter is removed.
The $\sigma_0$ values, also plotted in Fig.~\ref{dispersion}, suggest that the intrinsic
$(B-I)$ spread of each mode is near
$\sigma_0(blue) \simeq 0.10$ and $\sigma_0(red) \simeq 0.16$.
These translate to $\sigma$[Fe/H] = 0.27 dex for the metal-poor clusters
and 0.43 dex for the metal-rich ones, assuming that the linear conversion
from color to metallicity is valid.

\section{Radial Color Distributions}

\begin{figure}
\plotone{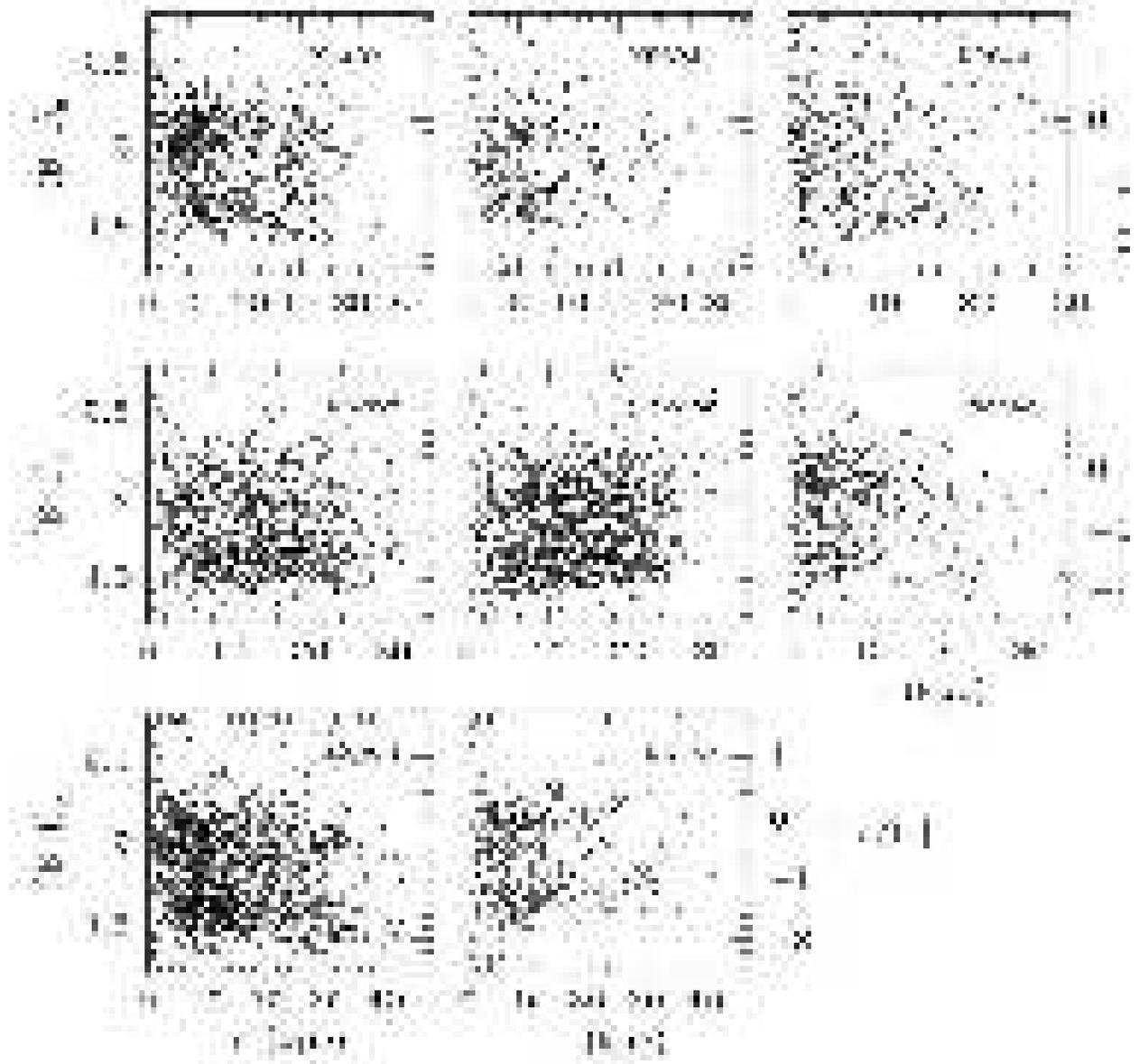}
\caption{$(B-I)_0$ versus radius for each of the BCG fields.
Here only the brighter objects are selected as described in the text,
to minimize field contamination from faint and predominantly blue
objects.  Color is converted to metallicity [Fe/H] following Eqn.~(5).
}
\label{birad8}
\end{figure}

Figure \ref{birad8} shows the dereddened cluster colors versus projected
galactocentric distance $r$; for convenience of comparison, we convert
the radial distances $r$ to kiloparsecs, and also show 
the [Fe/H] scale on the right-hand edge of each plot.
We find little or no significant trend of color with radius (thus
by assumption metallicity) {\sl within} either population. Any systematic
color gradient with galactocentric distance must then be the straightforward
result of a changing ratio of blue-to-red relative numbers with radius.
This latter effect was shown to exist for the GCSs in 
NGC 4472 by \citet{gei96} and has been found in other systems since
\citep[e.g.][]{harris98,dirsch03,rhode04}.

\begin{figure}
\plotone{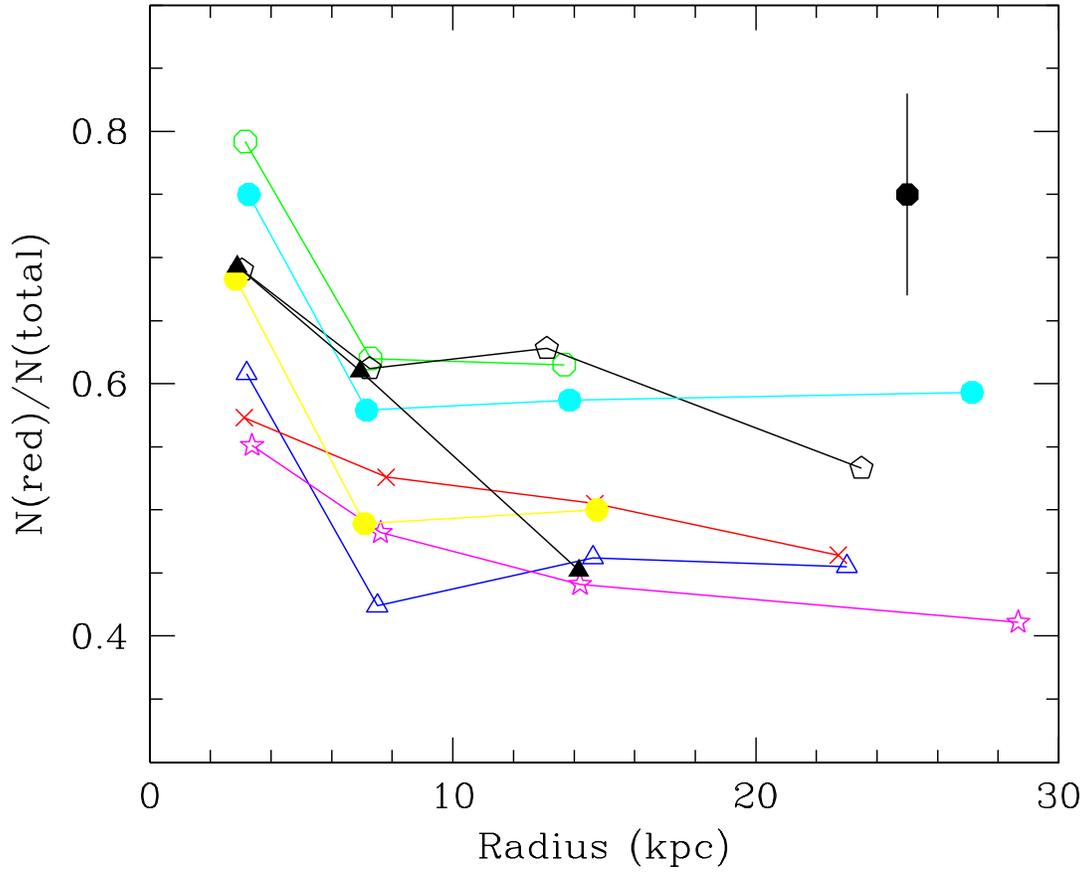}
\caption{Ratio of red clusters to total number of clusters, with data
taken from Table 3.  A typical error bar for the datapoints is
shown at upper right.  {\sl Open circles:}  NGC 1407.
{\sl Open triangles:} NGC 3258.  {\sl Crosses:}  NGC 3268.
{\sl Open pentagons:}  NGC 3348.  {\sl Open stars:}  NGC 4696.
{\sl Solid circles:}  NGC 5322 (lower) and NGC 5557 (upper).  
{\sl Solid triangles:} NGC 7049. }
\label{redbluetrend}
\end{figure}

As a simple indicator of the trend,
in Table \ref{bluered}, we show the progressive change in the ratio
$N(red)/N(total)$ with projected galactocentric radius for all eight systems.
These are listed in four radial bins,
$0 - 5$ kpc, $5 - 10$ kpc, $10 - 20$ kpc, and $> 20$ kpc.
For each bin, we list the total
number of clusters, then the ratio $N(red)/N(total)$.
To minimize field contamination when counting probable clusters, 
we simply exclude objects bluer than $(B-I)_0 = 1.3$,
redder than 2.3, or fainter than $M_I = -9.0$, and we 
define $(B-I)_0 = 1.8$ as the dividing line between the two populations.
The trends are plotted in Figure \ref{redbluetrend}.

All the GCSs show evidence for population gradients.  The largest part 
of the effect comes from 
within $R_{gc} = 5$ kpc, where the red clusters are always
more dominant, as would be expected if they belong to later formation
epochs when enriched gas had already concentrated deeper into the potential
well of the major galaxy.  In all our BCGs, the red clusters make up more than
half the population within 5 kpc, and can make up as much as 80\% in
the most extreme case.  Outward of 5 kpc, however, little or no gradient
shows up for any of the systems.  In 
a later paper, we will explore the radial distributions and
total populations in more detail.  

The total numbers of red vs. blue clusters 
that we observe in our target systems 
differ quite a lot from one galaxy to the next.  
The ratio $N(blue)/N(red)$ ranges
from $\sim 0.6$ in NGC 1407 up to $\sim 1.5$ in NGC 3258.
But at least part of this range may be due to the fact that the galaxies
are at different distances, so that we do not sample the same true range
in galactocentric radius $r_{gc}$ in each one.  For the closer systems we are
seeing only the inner and middle regions of the halo, and if the red
clusters are more centrally concentrated we will see proportionally more
of them than in an intrinsically similar but more distant galaxy.
In Table \ref{bluered}, we show the average $f(red)$ in the last line
for the various radial bins.  The values of $f(red)$ for the individual
galaxies have internal uncertainties that are only scarcely larger than
the rms scatter among the galaxies, indicating that 
the differences are not highly significant.
We note, however, that if the dominant formation mechanism for these BCGs is hierarchical
merging, then in a rough sense we would expect more 
galaxy-to-galaxy variance among
the red-cluster population, since they arise from a small number of
major mergers in the late stages of formation and thus are subject
to potentially large stochastic variations \citep{beasley03}.  By contrast, the blue
clusters are formed from a {\sl large} number of {\sl small} formation
events within the pregalactic dwarf population and thus their relative
numbers should be more consistent from one BCG to the next.

\section{Discussion}

The underlying cause for the bimodal nature of the globular cluster MDF
is still something of a mystery.  Ideas have been suggested involving
preferentially early or quick formation of globular clusters in the very
first stages of galaxy formation, when few stars had yet formed
and the gas was in the form
of numerous very low-metallicity, dwarf-sized clouds \citep{hh02,beasley02}.
This first round of formation would have produced many low-metallicity
clusters, but because it was happening in many relatively small potential
wells, much of the remaining gas could have been ejected by stellar winds
and supernovae, preventing it from being incorporated into stars till
much later.
However, these ideas have not yet developed into a full quantitative picture.
One important concern is that recent simulations \citep[e.g.,][]{mac99,fra04}
suggest dwarfs with more than $\sim 10^9 M_{\odot}$ of gas
would have suffered relatively little mass loss for plausible supernova rates.
Thus if early cluster formation went on in potential wells at least this
massive, there appear to be no reason why clusters could not have
continued to form into intermediate and high metallicity regimes.

In addition, it seems to be important that the same bimodality phenomenon now regarded to
be normal for globular clusters does {\sl not} show up in the
halo and bulge {\sl field stars} of E galaxies, the MDFs for
which are broad, unimodal, and contain relatively few metal-poor 
stars \citep{hh02}.  An alternate way to state this
dichotomy is that the specific frequency of the stellar population,
defined as the number of clusters per unit galaxy light within a given metallicity range, 
needs to be typically at least 
$\sim 5$ times higher for the metal-poor regime ([Fe/H] $< -1$) than for the metal-rich
regime [Fe/H] $> -1$ \citep{h01,harris02,forte05}. At face value, the evidence
suggests that large galaxies were very much more efficient
at producing high-mass star clusters at early times when the gas fraction
was extremely high and the metallicity extremely low, than they were later on
during the major field-star formation stages.

Conversely, it might be selective destruction rather 
than formation mechanisms that is responsible for the specific frequency difference. 
Perhaps the low-metallicity star clusters that form in dwarf 
galaxies are more likely to survive destruction during their 
infancy than their high-metallicity counterparts in larger 
galaxies \citep[e.g.][]{whit05}.  This effect might be especially relevant 
for clusters that form at the centers of dwarf galaxies \citep[e.g.][]{lotz04},
where the tidal stress is low and where they might be able to build up
to very large masses.  
The fundamental characteristic of bimodality (the rather distinct gap between
metal-poor and metal-rich subpopulations),
however, eludes satisfactory explanation as yet.

As we have done in the preceding discussion, we adopt here the ``default'' view 
that the cluster colors are indicators of heavy-element abundance rather than
age.  For clusters older than $\sim 5$ Gy, broadband 
optical colors are sensitive primarily
to metallicity and only secondarily to mean age.  
For comparison, it is worth noting that the metal-poor 
clusters in the Milky Way show no 
systematic differences in age larger than $\sim 1$ Gy
\citep[e.g.][]{ros99}, and the metal-rich clusters may be only
$\sim 2$ Gy younger in the mean.
However, the red-cluster population clearly exhibits a much larger
internal dispersion in the giant ellipticals than in the Milky Way,
perhaps reflecting the series of major mergers and starbursts that
the galaxy was built from in a hierarchical-merging scenario
\citep[e.g.,][]{beasley02,beasley03}.  
Under these conditions, a cluster-to-cluster age range of a few Gy
would be more likely {\sl a priori}.
Globular cluster age ranges at this level 
within selected giant E galaxies have recently been claimed through
spectral-index analysis, although the 
majority of the clusters still appear to be 10 Gy old or more 
\citep[e.g.,][]{beasley00,forbes01,jordan02,puz03,larsen03,hempel03,peng04}.
Even so, an age scatter of just a few Gy 
would generate a color spread of 0.1 mag or less,
not enough to match the intrinsic scatter in color that
we find in either the blue or red modes.

An excellent recent overview on globular cluster formation models is given
by \citet{krav05}.  Three major contributors to the GCS populations within
large galaxies conventionally include (a) early {\sl in situ} formation 
\citep{hp94,forbes97}, (b) major mergers that are sufficiently gas-rich to
form large numbers of new clusters \citep[][among many others]{whit99,zepf99}, 
and (c) ongoing accretion of gas-free satellites \citep{cot00}.
In certain circumstances, it may be possible to isolate
one or the other of these as the  dominant
mechanism. For example, ``field'' elliptical galaxies with low GCS specific
frequencies $S_N < 2$ and a wide mixture of cluster ages are excellent 
candidates to have formed from a small number of major mergers 
of disk galaxies \citep{h01}.
In the Milky Way, late mergers are not important, but
accretion of small, low-metallicity satellites may have been responsible
for building much of the halo \citep{cot00}.  

For the giant BCGs, we are more likely to
be looking at the combined result of all three processes in their full
complexity, making it more difficult to separate out their effects.  For 
this reason, in the following discussion
we adopt the viewpoint of a 
hierarchical-merging scheme arising from standard CDM cosmologies.
In this framework we essentially subsume all three individual processes into 
one (albeit complex) sequence:  at very early times, 
many small clouds within a large dark-matter potential well
merge progressively and rapidly into a ``seed'' giant elliptical; 
later on, a few major mergers may take place; then finally, a 
low-level process of satellite accretion continues to the present time.

Within this hierarchical-merging context,
the blue, metal-poor clusters are regarded to have
formed first within the original population of pregalactic gas clouds (the
so-called ``quiescent mode'', before truly big mergers were occurring). 
A similar view has frequently been argued on other grounds 
\citep{hp94,burg01,hh02,beasley02,krav05}.
The red, metal-rich clusters then formed during the later mergers that built
the main body of a giant galaxy, the latter
stages of which may still be continuing today.  Some major pieces of observational
evidence which are strongly consistent with this general view are:
\begin{itemize}
\item{} The MDF of the {\sl field stars} matches up well with the MDF of
the {\sl metal-rich} globular clusters \citep{hh02,harris02,forte05}, as expected 
if they formed in lockstep during the sequence of starbursts generated by mergers.
\item{} In most giant ellipticals including BCGs, the metal-rich clusters form
a more centrally concentrated radial distribution around their parent galaxy than
do the metal-poor clusters.  In turn, the red-cluster distribution is usually found
to match up well with the galaxy halo light profile as a whole, again consistent with 
their contemporaneous formation.
\item{} The globular clusters in dwarf elliptical galaxies are almost entirely
of low metallicity, consistent with their having evolved directly from protogalactic
gas clouds that experienced few or no mergers \citep{dur96a,lotz04}.
\item{} The mean colors of both the metal-rich and metal-poor modes increase
weakly with parent galaxy luminosity, indicating that the total size of its potential
well influenced the degree of heavy-element enrichment from the 
start \citep[e.g.][]{strader04,lotz04}.
\end{itemize}

Our new data provide additional constraints on the enrichment history
of the globular clusters.  
If the formal double-Gaussian fits (Fig.~\ref{zscale})
are on the right track, then we have a series of new observational
features to deal with:
\begin{itemize}
\item{} The formation of the low-metallicity
clusters (presumably the oldest ones)
must apparently allow for higher metallicity at higher cluster mass,
at least in some major galaxies such as BCGs;  
\item{} Contrarily, for the red (metal-rich) cluster sequence, there is no
trend in mean metallicity with luminosity;
\item{} At all luminosities, the cluster-to-cluster spread in
heavy-element abundance $\sigma_{log Z}$ is quite significant, and
is larger and more variable from galaxy to galaxy for the red
clusters than for the blue ones.
\end{itemize}

A plausible interpretation for the first item seems to us to be connected
with an idea previously stated in the literature, that {\sl on the average}, higher-mass
clusters require more massive reservoirs of gas out of which to form.
The typical observational
scaling ratio is that the host gas cloud is $\sim10^3$ times more massive than
any single average star cluster that forms within it \citep{hp94}.  The smaller
globular clusters -- those with masses
$< 10^5 M_{\odot}$ -- could then typically have formed within 
pregalactic clouds containing only $\sim 10^7-10^8 M_{\odot}$ of gas, 
while the very most massive globular clusters known
(those with $M > 10^7 M_{\odot}$) would have required reservoirs of gas
as high as $10^{9} - 10^{10} M_{\odot}$.  The potential wells within these bigger
clouds would be massive enough to hold on
to their gas during the first violent rounds of star formation and thus 
partially self-enrich shortly before
the first massive star clusters appeared, allowing them to build up to
higher metallicity.  At the same time, these most massive host dwarfs could
have escaped the ``gap'' or interruption at intermediate cluster metallicities
that affected the lower-mass clusters almost universally.
Clearly, however, some crucial elements of this descriptive story are still missing.

Further evidence that the metallicities of the massive clusters are intimately
tied to their parent cloud masses may come from two other scaling relations.
For dwarf galaxies, the mean metallicity is observed to vary with total stellar mass 
$M_{\star}$ as
\citep{dek03}
\begin{equation}
{Z \over Z_{\odot}}(dwarf) \, \simeq \, 4 \times 10^{-5} M_{\star}^{0.40}
\end{equation}
where $M_{\star}$ is in Solar units.  
However, it is the dark matter in
the dwarf that dominates its potential well, and an observationally based scaling of
the ratio of dark matter to stellar mass is \citep{per96}
\begin{equation}
{M_{DM} \over M_{\star} } \, \simeq \, 34.7 M_{\star,7}^{-0.29}
\end{equation}
where here $M_{\star,7}$ is in units of $10^7 M_{\odot}$.  Combining Eqs.~(9) 
and (10), we obtain
\begin{equation}
{Z \over Z_{\odot}}(dwarf) \, = \, 3.9 \times 10^{-7} M_{DM}^{0.56}.
\end{equation}
This latter relation has a similar slope to the one suggested above for
the metal-poor clusters.  At this stage, we view this comparison as little
more than suggestive.  However, the
implication for our purposes is that the most massive star clusters
can end up with heavy-element abundances
determined directly by the dark-matter potential wells of the pregalactic
clouds that they formed in, since $M_{DM}$ would determine the metallicity
of the ambient gas from which they formed.  
Notably, \citet{lotz04} find, for the globular cluster populations in dwarf E galaxies,
that the mean metallicity $Z(GC)$ of the clusters scales with the luminosity
of the host galaxy as $Z(GC) \sim L_{dE}^{0.2}$.  This is a 
slightly shallower trend than the
ones for the dwarfs themselves, but one which also argues that
the enrichment of a given cluster increases with the mass of its host environment.

As mentioned above, current numerical simulations 
\citep[see, e.g., the extensive discussions of][]{mac99,fra04} indicate that 
dwarfs with gas masses less than $10^8 M_{\odot}$ have heavy-element
ejection efficiencies approaching 100\% for a range of plausible supernova
rates and distributions.  However, model dwarfs with masses above
$10^9 M_{\odot}$ successfully retain their 
heavy elements except under extremely high starburst conditions.  
It is the dwarfs in this upper range that should be capable of
building globular clusters with masses well above $10^6 M_{\odot}$.

The cluster-to-cluster {\sl scatter} of metallicity at a given mass could
well have been generated by the random nature of the starbursts within their
host dwarfs, since the metal ejection efficiency depends on the detailed locations of
the individual supernovae as well as their ignition rates (cf.~the references cited above).

Lastly, within the same hierarchical-merging picture, why would the metal-richer cluster
sequence not show the same kind of mass/metallicity relation as the blue clusters?
We suggest that the explanation may lie in the relative epoch of their
formation.  These objects would have formed in the bigger, later mergers and starbursts that built
the main stellar population of the galaxy.  Each of these major bursts
could have assembled gas clouds (GMCs and GMC complexes) of a very wide range of sizes and
masses, within which star clusters of a similarly broad range could form,
such as we see happening in contemporary mergers
\citep{whit99,zhang01,wilson03}.  What distinguishes these later stages from
the earlier initial stage is that the large-scale potential
well of the new BCG has now formed, and gas cooling times are shorter at the
higher metallicities,
allowing newly enriched gas to be retained throughout the body of the galaxy.
The combination of these factors may then have allowed this second major
group of globular clusters to form along with the field stars 
in a common MDF that is extremely broad \citep{hh02}, and with no net
mass-metallicity relation.  

As mentioned earlier, 
we believe that another important characteristic of the metal-richer population 
is that both its {\sl mean metallicity} and {\sl internal metallicity
dispersion} are less uniform from one galaxy to the next
than those of the metal-poor population.  These
can be seen from the plots of $\langle B-I \rangle$ and $\sigma(B-I)$
(Fig~\ref{fits}), where the red population displays variations twice as large as
for the blue.
We suggest from these comparisons that the red population differs 
internally in both metallicity and age to a degree 
larger than in the bluer mode.  As noted above,
such differences would be expected if they were primarily formed in
the last few major mergers and thus subject to larger stochastic variations.

In this general scheme, the metal-richer cluster formation mode can
be thought of as the ``normal'' one
in the sense that these clusters match well with the MDF of the halo stars in 
the galaxy. It is the
metal-poor ones that are anomalous because of their very high specific
frequency.  Further quantitative models tuned specifically to following
globular cluster formation will need to be pursued in order to understand
exactly what was special about that first, remarkably efficient mode.

\section{Summary}

We present new HST ACS/WFC photometry in $(I, B-I)$ for globular cluster systems
in eight giant ellipticals, all of which are centrally dominant 
objects (BCGs) in clusters of galaxies.  In all cases the photometry
reaches to limits at or beyond the turnover point in the globular cluster
luminosity function, and clearly resolves the internal spread in colors
of the clusters.

All of the $(B-I)$ distributions are clearly bimodal, strongly supporting
the view built up in recent years that globular clusters in major galaxies come in two
distinctive types: metal-rich and metal-poor.  With the use of a metallicity-sensitive
color index and high-quality photometry, the structure of the metallicity
distribution function (MDF) is more clearly revealed here than in all but the best
previous studies.  

We find that in several of our target galaxies, the blue and red modes
effectively overlap or merge at high luminosity, specifically in the
range $M_I < -10.5$ (corresponding to cluster masses higher than $10^6 M_{\odot}$).  
At these high-luminosity, high-mass levels, extremely metal-poor clusters become
much rarer, and intermediate-metallicity clusters much more common
than at the lower levels where bimodality is much more sharply evident.
This effect is a result of a correlation between luminosity and color that sets
in within the metal-poor cluster sequence, shifting it progressively redward
towards higher luminosity and causing it to overlap with the red cluster sequence.
Furthermore, the larger the total measured cluster
population contained in the galaxy, the more clearly this feature emerges.
Comparisons with other published studies suggest to us that this phenomenon
may be particularly associated with these giant BCGs.
A compelling interpretation is not yet obvious, but if we make the plausible assumption
that more massive clusters on average must form within proportionally
larger parent gas clouds, then our data suggest to us that the most massive globular clusters
formed within parent clouds that were massive enough in turn
to hold on to more of their gas and reach higher levels
of pre-enrichment.  The recent literature suggests that the enrichment effect
will set in for dwarfs of gas masses
$10^9 M_{\odot}$ or more, large enough to produce globular cluster masses
of $> 10^6 M_{\odot}$.

We find that both the intrinsic color dispersion and mean color
of the red-cluster population differ more from galaxy to galaxy than the
blue-cluster population.  We do not see any pattern
to these differences; they are not obviously correlated with galaxy
luminosity, total measured cluster population, blue-to-red relative numbers
of clusters, or the richness of the galaxy cluster they are in.
Within the context of standard hierarchical-merging galaxy formation models,
we suggest that these very noticeable differences in the red-cluster population
may be due to stochastic variations in the last few major mergers that
built the galaxy.  By contrast, the blue-cluster population was built 
through a large number of smaller-scale events in the pregalactic cloud
population and thus is more nearly universal in form.

Finally, we find clear evidence for population gradients in all
the BCGs in our sample.  Within $\sim 5$ kpc of galaxy center, every galaxy
contains a distinctly higher proportion of red, metal-rich clusters.  Beyond
10 kpc, however, the residual color gradient becomes much smaller.

The two major and long-standing issues that our data reinforce are (1) the basic
bimodality property of the globular cluster MDF, which still lacks a
convincing physical explanation, and (2) the large numbers of metal-poor
clusters relative to the small numbers of metal-poor field halo stars that these
giant galaxies possess.  If the hierarchical-merging models are essentially correct,
in their earliest star-forming era the pregalactic clouds went through a short
period of producing massive star clusters at high efficiency.  The later production
of the redder, metal-richer clusters through major mergers and starbursts can now
be thought of as a familiar and reasonably well understood process, since modern-day
gas-rich mergers are seen producing these kinds of clusters actively.  It is the
earlier, pregalactic era that produced roughly the same total number
of massive clusters as all the later epochs combined, and that is still poorly understood.

In later papers, we will present further MDF data for five more BCGs, and
discuss their globular cluster luminosity distributions, specific frequencies,
and spatial structures.

\acknowledgments
This work was supported by the Natural Sciences and Engineering
Research Council of Canada through research grants to WEH and DAH, and
also by NASA grant GO-09427.01-A to BCW.  


\clearpage


\begin{deluxetable}{cccccccc}
\tabletypesize{\footnotesize}
\tablecaption{Brightest Cluster Galaxies Imaged with the ACS \label{basicdata}}
\tablewidth{0pt}
\tablehead{
\colhead{NGC} & \colhead{Cluster} & \colhead{Redshift} & 
\colhead{$M_V^T$} & \colhead{$E(B-I)$} & \colhead{$(m-M)_I$} & 
\colhead{$B$ Exposures} & \colhead{$I$ Exposures} \\
& \colhead{or Group} & \colhead{(km s$^{-1}$)} & & & \colhead{($M_I<-8.4$)} & 
\colhead{(sec)} & \colhead{(sec)} \\
}

\startdata
1407 & Eridanus & 1627 & $-22.35$ & 0.16 & 31.96 &  $2\times 750$ & $2 \times 340$ \\
5322 & CfA 122 & 1916 & $-22.01$ & 0.03 & 32.22 &  $3\times 1130$ & $2\times 410$ \\
7049 & N7049 & 1977 & $-21.76$ & 0.12 & 32.36 &  $6\times 580$ & $3\times 400$ \\
3348 & CfA 69 &  2837 & $-22.13$ & 0.17 & 33.18  & $6 \times 1200$ & $4\times 530$ \\
3258 & Antlia & 3129 & $-21.87$ & 0.20 & 33.23 &  $4 \times 1340$ & $4\times 570$ \\
3268 & Antlia & 3084 & $-21.96$ & 0.24 & 33.27 &  $4\times 1340$ & $4\times 570$ \\
4696 & Cen30 & 2926 & $-23.31$ & 0.23 & 33.29 &  $4\times 1360$ & $4\times 580$ \\
5557 & CfA 141 & 3213 & $-22.32$ & 0.01 & 33.31 &  $2\times 1280, 2\times 1350$ & $4\times 600$ \\

\enddata

\end{deluxetable}

\begin{deluxetable}{cccc}
\tabletypesize{\footnotesize}
\tablecaption{Measured Populations and Field Contamination Levels\label{field}}
\tablewidth{0pt}
\tablehead{
\colhead{NGC} & \colhead{$N_c$} & \colhead{$N_f$} & \colhead{$N_f/N_c$} \\
}

\startdata
1407 &   402 & 2.5  & 0.01 \\
5322 &   126 & 2.5 & 0.02 \\
7049 &   196 & 22 & 0.11 \\
3348 &   374 & 15 & 0.04 \\
3258 &   885 & 31 & 0.04 \\
3268 &   780 & 29 & 0.04 \\
4696 &  1587 & 116 & 0.07\\
5557 &   357 & 18 & 0.05 \\

\enddata

\end{deluxetable}

\begin{deluxetable}{ccccccc}
\tablecaption{Bimodal Fitting Parameters \label{bimodal}}
\tablewidth{0pt}
\tablehead{
\colhead{NGC} & \colhead{$N_b/N_r$} & \colhead{$\langle B-I \rangle_0$} & 
\colhead{$\sigma_{B-I}$} & \colhead{$\langle B-I \rangle_0$} & \colhead{$\sigma_{B-I}$} &
\colhead{$\Delta \langle B-I \rangle$} \\
& & \colhead{(blue)} & \colhead{(blue)} & \colhead{(red)} & \colhead{(red)} & \colhead{(red$-$blue)} \\
}

\startdata
1407 & 0.60 & $1.63\pm0.02$ & $0.12\pm0.02$ & $2.07\pm0.01$ & $0.13\pm0.01$ &$0.44\pm0.03$ \\
5322 & 1.10 & $1.67\pm0.01$ & $0.07\pm0.01$ & $2.01\pm0.02$ & $0.09\pm0.01$ &$0.46\pm0.02$\\
7049 & 1.25 & $1.67\pm0.04$ & $0.12\pm0.02$ & $2.14\pm0.06$ & $0.21\pm0.14$ &$0.47\pm0.07$\\
3348 & 0.70 & $1.60\pm0.01$ & $0.07\pm0.01$ & $2.07\pm0.02$ & $0.17\pm0.02$ &$0.47\pm0.02$\\
3258 & 1.49 & $1.61\pm0.01$ & $0.10\pm0.01$ & $2.01\pm0.03$ & $0.20\pm0.04$ &$0.40\pm0.03$ \\
3268 & 1.10 & $1.61\pm0.01$ & $0.10\pm0.01$ & $2.01\pm0.03$ & $0.20\pm0.04$ &$0.40\pm0.03$ \\
4696 & 1.31 & $1.67\pm0.01$ & $0.11\pm0.01$ & $2.09\pm0.01$ & $0.16\pm0.01$ &$0.42\pm0.02$\\
5557 & 0.91 & $1.66\pm0.02$ & $0.10\pm0.01$ & $2.10\pm0.02$ & $0.17\pm0.03$ &$0.44\pm0.03$\\
\\
BCG Mean & 1.06 & 1.64 (0.03) & 0.10 (0.02) & 2.06 (0.05) & 0.17 (0.04) & 0.44 (0.03) \\
\\
Milky Way & 2.34 & $1.57\pm0.02$ & $0.13\pm0.02$ & $1.95\pm0.02$ & $0.06\pm0.02$ &$0.38\pm0.03$\\

\enddata

\end{deluxetable}

\begin{deluxetable}{ccccccc}
\tabletypesize{\footnotesize}
\tablecaption{Gaussian Fits in Luminosity Bins for Individual Galaxies\label{brightKMM}}
\tablewidth{0pt}
\tablehead{
\colhead{NGC} & \colhead{$M_I = (-13.5, -10.5)$} & \colhead{$N$} & 
\colhead{$M_I = (-9.8, -9.5)$} & \colhead{$N$} & \colhead{$M_I = (-10.5, -9.0)$} & \colhead{$N$} \\
}

\startdata
1407 & 0.128 & 79 & $6.2 \times 10^{-7}$ & 85 & $3.0 \times 10^{-17}$ & 401 \\
5322 & 0.12  & 13 &                   -- & -- & $1.1 \times 10^{-7}$ & 102 \\
7049 & 0.161 & 45 & 0.016                & 39 & $2.3 \times 10^{-6}$ & 155 \\
3348 & 0.238 & 66 & $1.8 \times 10^{-6}$ & 67 & $1.6 \times 10^{-11}$ & 319 \\
3258 & 0.038 & 167 & $5.7 \times 10^{-9}$ & 165 & $3.6 \times 10^{-19}$ & 735 \\
3268 & --    & 161 & $9.5 \times 10^{-6}$ & 136 & $4.2 \times 10^{-15}$ & 632 \\
4696 & 0.0015 & 342 & $1.3 \times 10^{-9}$ & 231 & $2.9 \times 10^{-31}$ & 1260 \\
5557 & 0.0024 & 67 & $0.016$              & 69 & $1.6 \times 10^{-10}$ & 304 \\

\enddata

\end{deluxetable}

\begin{deluxetable}{ccccccccc}
\tablecaption{Red Cluster Population Ratios Versus Galactocentric Distance \label{bluered}}
\tablewidth{0pt}
\tablehead{
\colhead{NGC} &  \colhead{$0 - 5$} & \colhead{$f(red)$} & \colhead{$5 - 10$} &
\colhead{$f(red)$} & \colhead{$10 - 20$} & \colhead{$f(red)$} &
\colhead{$> 20$} & \colhead{$f(red)$} \\
&  \colhead{kpc} & & \colhead{kpc}  & & \colhead{kpc} & & \colhead{kpc}  \\
}

\startdata
1407 &  173 & $0.79\pm0.09$ & 163 & $0.62\pm0.08$ &
130 & $0.62\pm0.09$ & 3 & $-$ \\

5322 & 63 & $0.68\pm0.14$ & 45 & $0.49\pm0.13$ & 
18 & $0.50\pm0.20$ & 0 & $-$ \\

7049 & 75 & $0.69\pm0.13$ & 82 & $0.61\pm0.11$ & 62
& $0.45\pm0.10$ & 7 & $-$ \\

3348 & 145 & $0.69\pm0.09$ & 116 & $0.61\pm0.09$ &
78 & $0.63\pm0.12$ & 30 & $0.53\pm0.17$ \\

3258 & 125 & $0.61\pm0.09$ & 224 & $0.42\pm0.05$ &
418 & $0.46\pm0.04$ & 110 & $0.46\pm0.08$ \\

3268 & 96 & $0.57\pm0.10$ & 196 & $0.53\pm0.06$ &
400 & $0.51\pm0.04$ & 84 & $0.46\pm0.09$ \\

4696 & 156 & $0.55\pm0.07$ & 386 & $0.48\pm0.04$ &
580 & $0.44\pm0.03$ & 555 & $0.41\pm0.03$ \\

5557 & 76 & $0.75\pm0.13$ & 107 & $0.58\pm0.09$ &
109 & $0.59\pm0.09$ & 54 & $0.59\pm0.13$  \\

\\
Mean & & $0.67\pm0.03$ & & $0.54\pm0.03$ & & $0.53\pm0.03$ & & $0.49\pm0.03$ \\

\enddata

\end{deluxetable}

\clearpage


\clearpage

\clearpage

\clearpage

\clearpage

\clearpage

\clearpage

\clearpage

\clearpage

\clearpage

\clearpage

\clearpage

\clearpage

\clearpage

\clearpage

\clearpage

\clearpage

\clearpage

\clearpage

\clearpage

\clearpage

\clearpage

\clearpage

\end{document}